\documentclass{article}
\usepackage{amsmath}
\usepackage{amsthm}
\usepackage{amsfonts}
\usepackage{amssymb}
\usepackage{amscd}
\usepackage[mathscr]{eucal}
\usepackage{mathtools}
\usepackage{fullpage}
\usepackage{color}
\usepackage{float}
\usepackage{graphicx}
\usepackage{textcomp}

\newcommand*{\myabstract}[1]{\begin{abstract} #1 \end{abstract} }
\begin{document}
\title{\bf On Backus average in modelling guided waves}
\author{
David R. Dalton%
\footnote{Department of Earth Sciences, Memorial University of Newfoundland,  Canada; 
{\tt dalton.nfld@gmail.com}}\,,
Thomas B. Meehan%
\footnote{Faculty of Engineering, Memorial University of Newfoundland,  Canada; 
{\tt tbm247@mun.ca}}\,,
Michael A. Slawinski%
\footnote{Department of Earth Sciences, 
Memorial University of Newfoundland, Canada; 
{\tt mslawins@mac.com}}}
\maketitle
\myabstract{%
We examine the Backus~\cite{backus} average of a stack of isotropic layers overlying an isotropic  halfspace to examine its applicability for the quasi-Rayleigh and Love wave dispersion curves, both of which apply to the same model.  
We compare these curves to values obtained for the stack of layers using the propagator matrix.
The Backus~\cite{backus} average is applicable only for thin layers or low frequencies.
This is true for both weakly inhomogeneous layers resulting in a weakly anisotropic medium and strongly inhomogeneous alternating layers resulting in a strongly anisotropic medium.
We also compare the strongly anisotropic and weakly anisotropic media, given by the Backus~\cite{backus} averages, to results obtained by the isotropic Voigt~\cite{voigt} averages of these media.
As expected, we find only a small difference between these results for weak anisotropy and a large difference for strong anisotropy.
We perform the Backus~\cite{backus} average for a stack of alternating transversely isotropic layers that is strongly inhomogeneous to evaluate the dispersion curves for the resulting medium.  
We compare these curves to values obtained using a propagator matrix for that stack of layers.
Again, there is a good match only for thin layers or low frequencies. 
Finally, we perform the Backus~\cite{backus} average for a stack of nonalternating transversely isotropic layers that is strongly inhomogeneous, and evaluate the quasi-Rayleigh wave dispersion curves for the resulting transversely isotropic medium.  
We compare these curves to values obtained using the propagator matrix for the stack of layers.
In this case, the Backus~\cite{backus} average performs less well, but---for the fundamental mode---remains adequate for low frequencies or thin layers.}
\section{Introduction}
This paper is an examination of the applicability of the Backus~\cite{backus} average to guided-wave-dispersion modelling.
We compare the dispersion curves of both Love and quasi-Rayleigh waves for the Backus~\cite{backus} average of a stack of layers to the dispersion curves for these layers using the propagator-matrix method.
The prefix distinguishes quasi-Rayleigh waves, which are guided waves, from classical Rayleigh waves, which propagate within a halfspace.
We examine the effects of strength of inhomogeneity, anisotropy and layer thickness.

The focus on examining both Love and quasi-Rayleigh waves is motivated by their existence in the same model.
This is a consequence of compatibility of their wave equations and boundary conditions, as discussed by Dalton et al.~\cite{dalton}.

Similar work---for Love waves in a stack of alternating isotropic layers---was done by Anderson~\cite{anderson62}, in which the author drew on the results of Postma~\cite{postma}.
Herein, we broaden this work to quasi-Rayleigh waves and, drawing on the work of Backus~\cite{backus}, to a nonalternating stack of isotropic layers.
We extend the scope to include Love waves for a stack of alternating transversely isotropic layers and quasi-Rayleigh waves for stacks of alternating and nonalternating transversely isotropic layers.

We begin this paper by providing background information for the Backus~\cite{backus} and Voigt~\cite{voigt} averages, as well as the Thomsen~\cite{thomsen} parameters.
The essence of this paper consists of numerical results and their discussion, where we consider the effects of strength of inhomogeneity, anisotropy and layer thickness on different modes and frequencies of the dispersion relations for the quasi-Rayleigh and Love waves.
\section{Background}
Backus~\cite{backus} shows that---for thicknesses of individual layers that are much less than the wavelength---waves travelling through parallel isotropic layers behave as if they were travelling through a single transversely isotropic medium.
An examination and extension of the Backus~\cite{backus} average, as well as its limitations, are discussed by Bos et al.~\cite{bos,BosX}.  

The parameters of this medium are (e.g., Slawinski~\cite[equations~(4.37)--(4.42)]{slawinski3}) 
\begin{equation}
\label{eq:Tue1}
c^{\overline{\rm TI}}_{1111}=\overline{\left(\frac{c_{1111}-2c_{2323}}{c_{1111}}\right)}^{\,2}
\,\,\,\overline{\left(\frac{1}{c_{1111}}\right)}^{\,-1}
+\overline{\left(\frac{4(c_{1111}-c_{2323})c_{2323}}{c_{1111}}\right)}
\,,
\end{equation}
\begin{equation}
\label{eq:Backus1133}
c^{\overline{\rm TI}}_{1133}=\overline{\left(\frac{c_{1111}-2c_{2323}}{c_{1111}}\right)}\,\,
\,\,\overline{\left(\frac{1}{c_{1111}}\right)}^{\,-1}
\,,
\end{equation}
\begin{equation}
\label{eq:Berry1}
c^{\overline{\rm TI}}_{1212}=\overline{c_{2323}}
\,,
\end{equation}
\begin{equation}
\label{eq:Berry2}
c^{\overline{\rm TI}}_{2323}=\overline{\left(\frac{1}{c_{2323}}\right)}^{\,-1}
\,,
\end{equation}
\begin{equation}
\label{eq:Tue2}
c^{\overline{\rm TI}}_{3333}=\overline{\left(\frac{1}{c_{1111}}\right)}^{\,-1}
\,,
\end{equation}
where $c_{ijk\ell}$ are the elasticity-tensor components for an isotropic Hookean solid and the overline indicates an average.
These expressions constitute a medium equivalent to a stack of layers, which we refer to as the Backus medium.  
In this paper, we use an arithmetic average, whose weight is the layer thickness, which we take to be the same for all averaged layers; for example,
\begin{equation}
\label{eq:weight}
c^{\overline{\rm TI}}_{1212}=\overline{c_{2323}}=\frac{1}{n}\sum_{i=1}^n\left(c_{2323}\right)_i\,,
\end{equation}
where $n$ is the number of layers.

Backus~\cite{backus} also shows that waves travelling through parallel transversely isotropic layers behave as if they were travelling through a single transversely isotropic medium.   
The parameters of such a medium are (e.g., Slawinski~\cite[equations~(4.54)]{slawinski3})
\begin{equation}
\label{eq:c1111}
c^{\overline{\rm TI}}_{1111}=
\overline{\left(c_{1111}-\frac{c_{1133}^2}{c_{3333}}\right)}
+\overline{\left(\frac{c_{1133}}{c_{3333}}\right)}^{\,2}
\,\,\,\overline{\left(\frac{1}{c_{3333}}\right)}^{\,-1}
\,,
\end{equation}
\begin{equation}
\label{eq:c1133}
c^{\overline{\rm TI}}_{1133}=\overline{\left(\frac{c_{1133}}{c_{3333}}\right)}
\,\,\,\overline{\left(\frac{1}{c_{3333}}\right)}^{\,-1}
\,,
\end{equation}
\begin{equation}
\label{eq:c11212}
c^{\overline{\rm TI}}_{1212}=\overline{c_{1212}}
\,,
\end{equation}
\begin{equation}
\label{eq:c2323}
c^{\overline{\rm TI}}_{2323}=\overline{\left(\frac{1}{c_{2323}}\right)}^{\,-1}
\,,
\end{equation}
\begin{equation}
\label{eq:c3333}
c^{\overline{\rm TI}}_{3333}=\overline{\left(\frac{1}{c_{3333}}\right)}^{\,-1}
\,,
\end{equation}
where $c_{ijk\ell}$ are the elasticity-tensor components of a transversely isotropic Hookean solid.
This result is also referred to as the Backus medium.
The parameters in expressions~(\ref{eq:Tue1})--(\ref{eq:Tue2}) and in expressions~(\ref{eq:c1111})--(\ref{eq:c3333}) are denoted by $c^{\overline{\rm TI}}_{ijk\ell}$\,.
However---even though the former are a special case of the latter and both share the same material symmetry---they correspond to distinct media, as shown by different expressions on the right-hand sides of the corresponding equations in systems (\ref{eq:Tue1})--(\ref{eq:Tue2}) and (\ref{eq:c1111})--(\ref{eq:c3333}).

To quantify the strength of anisotropy of transversely isotropic media, we invoke the three Thomsen~\cite{thomsen} parameters that are zero for isotropy and have absolute values much less than one for weak anisotropy,
\begin{equation}
\label{eq:gamma}
\gamma:=\frac{c^{\overline{\rm TI}}_{1212}-c^{\overline{\rm TI}}_{2323}}{2c^{\overline{\rm TI}}_{2323}}
\,,
\end{equation}
\begin{equation}
\label{eq:delta}
\delta:=\frac{\left(c^{\overline{\rm TI}}_{1133}+c^{\overline{\rm TI}}_{2323}\right)^2-\left(c^{\overline{\rm TI}}_{3333}-c^{\overline{\rm TI}}_{2323}\right)^2}{2c^{\overline{\rm TI}}_{3333}\left(c^{\overline{\rm TI}}_{3333}-c^{\overline{\rm TI}}_{2323}\right)}
\,,
\end{equation}
\begin{equation}
\label{eq:epsilon}
\epsilon:=\frac{c^{\overline{\rm TI}}_{1111}-c^{\overline{\rm TI}}_{3333}}{2c^{\overline{\rm TI}}_{3333}}
\,.
\end{equation}

To examine the effects of anisotropy, we study dispersion curves for the closest isotropic counterpart, as formulated by Voigt~\cite{voigt}; this formulation is an isotropic case of the Gazis et al.~\cite{gazis} average.
The two elasticity parameters of the isotropic counterpart of a Backus medium are (Slawinski~\cite[equations~(4.77) and (4.78)]{slawinski3})
\begin{equation}
\label{eq:ISO1111TI}
c^{\overline{\rm iso}}_{1111}=\frac{1}{15}\left(8c^{\overline{\rm TI}}_{1111}+4c^{\overline{\rm TI}}_{1133}+8c^{\overline{\rm TI}}_{2323}+3c^{\overline{\rm TI}}_{3333}\right)
\end{equation}
and
\begin{equation}
\label{eq:ISO2323TI}
c^{\overline{\rm iso}}_{2323}=\frac{1}{15}\left(c^{\overline{\rm TI}}_{1111}-2c^{\overline{\rm TI}}_{1133}+5c^{\overline{\rm TI}}_{1212}+6c^{\overline{\rm TI}}_{2323}+c^{\overline{\rm TI}}_{3333}\right)
\,;
\end{equation}
henceforth, this result is referred to as the Voigt medium. 
\section{Isotropic layers}
\label{sec:IsoLay}
Our study of quasi-Rayleigh and Love waves is conducted by examining their dispersion relations.
For a Voigt medium isotropic layer of thickness $Z$ with density $\rho^u$\,, $S$-wave speed $\beta^u=\sqrt{c_{2323}^u/\rho^u}$\,, and $P$-wave speed $\alpha^u=\sqrt{c_{1111}^u/\rho^u}$ overlying an isotropic halfspace with density $\rho^d$\,, $S$-wave speed $\beta^d=\sqrt{c_{2323}^d/\rho^d}$\,, and $P$-wave speed $\alpha^d=\sqrt{c_{1111}^d/\rho^d}$\,, the dispersion relations for Love and quasi-Rayleigh waves are given in equations~(2) and~(17), respectively, of Dalton et al.~\cite{dalton}. Herein, these relations are coded in Mathematica\textsuperscript{\textregistered} and the dispersion curves are plotted as zero contours of the respective dispersion relations, which are purely real.

An insight into the dispersion relations based on the Backus~\cite{backus} average is provided by comparing their curves to the dispersion curves computed for a stack of layers from which the Backus medium is obtained by averaging their properties.
For a stack of isotropic layers overlying an isotropic halfspace, the dispersion relations for Love and quasi-Rayleigh waves are based on a propagator matrix, specifically, on the delta-matrix solution reviewed in Buchen and Ben-Hador~\cite{buchen}.
These relations are coded in Python\textsuperscript{\textregistered}.

\begin{figure}
\begin{center}
\includegraphics[scale=0.55]{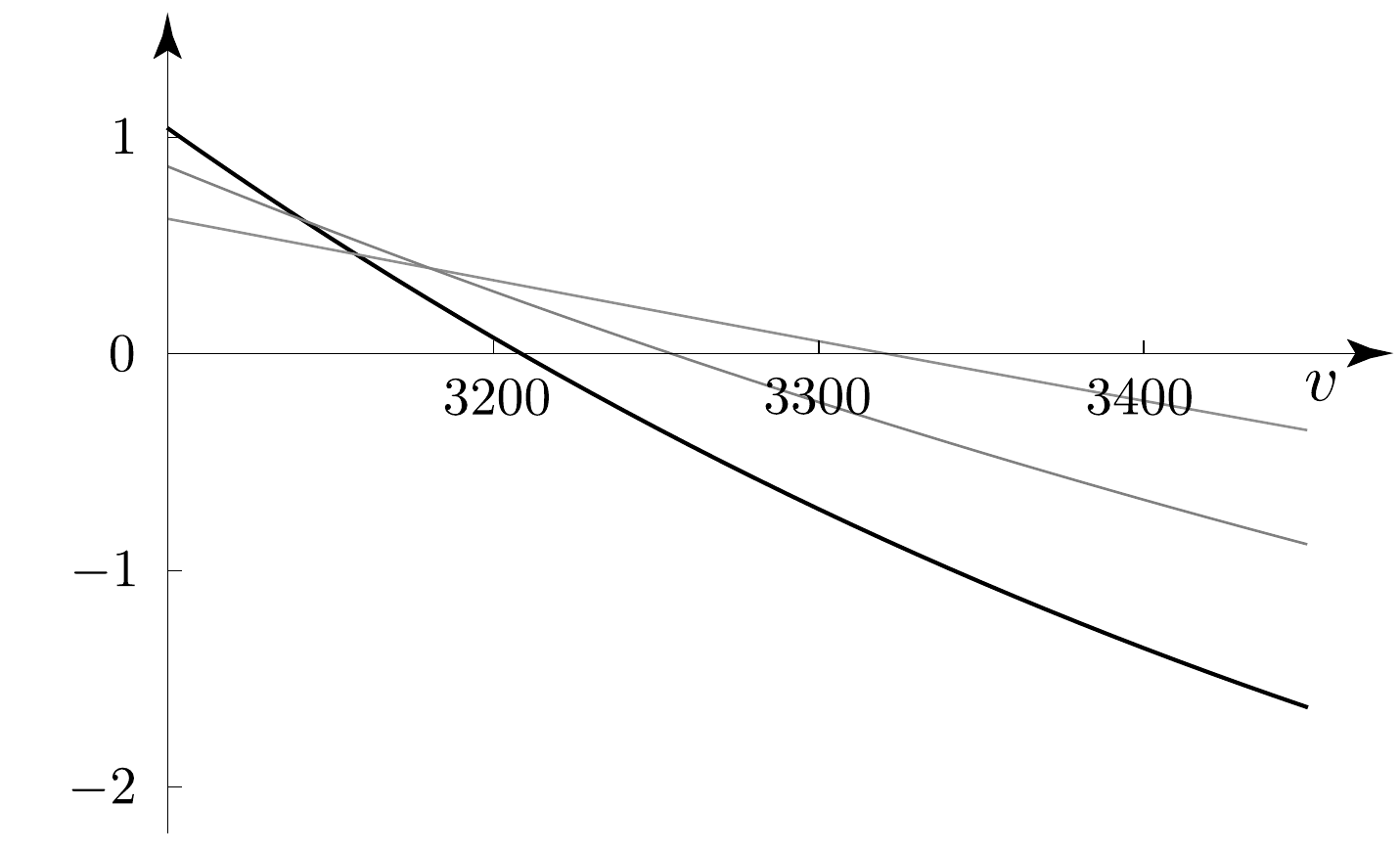}
\includegraphics[scale=0.55]{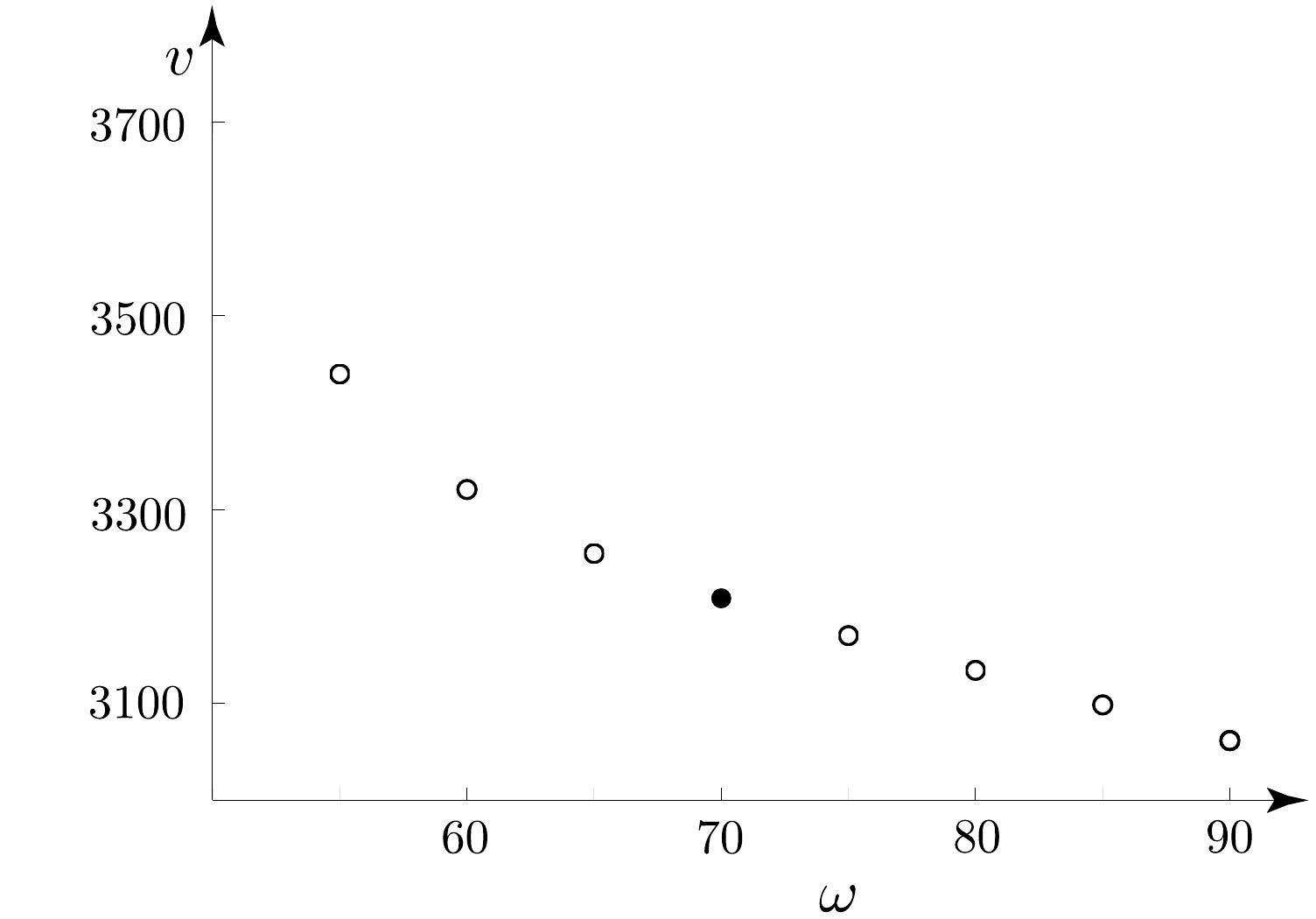}
\end{center}
\caption{\small{Dispersion curve obtained by solving for~$v$  the dispersion relations with a given value of $\omega$}}
\label{fig:Thomas}
\end{figure}

In contrast to the Backus~\cite{backus} average, the propagator matrix allows us to combine information about all layers while retaining their individual properties.
To illustrate the root-finding process, in the left-hand plot of Figure~\ref{fig:Thomas}, we see the values of the determinant for the dispersion relation as a function of speed.
Each curve corresponds to a distinct value of~$\omega$\,; the black curve corresponds to $\omega=70\,{\rm s}^{-1}$\,.
For each curve, the solution corresponds to $v$ for which the determinant is equal to zero.
The right-hand plot is a dispersion curve, where each point, $(\omega,\,v)$\,, is a solution from the left-hand plot; the black dot at $\omega=70\,{\rm s}^{-1}$ corresponds to the zero intercept of the black curve.
More details are included in Meehan~\cite{meehancode,meehanarxiv}.

\subsection{Weak inhomogeneity}
Let us consider a weakly inhomogeneous stack of isotropic layers, whose elasticity parameters and corresponding velocities are given in Table~\ref{table:BackusAverage}.
\begin{table}[h]
\begin{center}
\begin{tabular}{|c||c|c|c|c|c|}
\hline
layer & $c_{1111}$ & $c_{2323}$ & $v_P$ & $v_S$ \\
\hline\hline
1 & 10.56 & 2.02 & 3.25 & 1.42 \\
2 & 20.52 & 4.45 & 4.53 & 2.11 \\
3 & 31.14 & 2.89 & 5.58 & 1.70 \\
4 & 14.82 & 2.62 & 3.85 & 1.62 \\
5 & 32.15 & 2.92 & 5.67 & 1.71 \\
6 & 16.00 & 2.56 & 4.00 & 1.60 \\
7 & 16.40 & 6.35 & 4.05 & 2.52 \\
8 & 18.06 & 4.33 & 4.25 & 2.08 \\
9 & 31.47 & 8.01 & 5.61 & 2.83 \\
10 & 17.31 & 3.76 & 4.16 & 1.94 \\
\hline
\end{tabular}
\end{center}
\caption{\small{Density-scaled elasticity parameters,~$\times 10^6\,{\rm m}^{2}\,{\rm s}^{-2}$\,, for a weakly inhomogeneous stack of isotropic layers, and the corresponding $P$-wave and $S$-wave velocities,~${\rm km\,s}^{-1}$ (Brisco~\cite{brisco} and Slawinski~\cite{slawinski3})}}
\label{table:BackusAverage}
\end{table}

Following expressions~(\ref{eq:Tue1})--(\ref{eq:Tue2}), we obtain
$c^{\overline{\rm TI}}_{1111}=18.84$\,,
$c^{\overline{\rm TI}}_{1133}=10.96$\,,
$c^{\overline{\rm TI}}_{1212}=3.99$\,,
$c^{\overline{\rm TI}}_{2323}=3.38$ and
$c^{\overline{\rm TI}}_{3333}=18.43$\,;
these values are to be multiplied by $10^6$\,, and their units are ${\rm m}^{2}\,{\rm s}^{-2}$\,.

The nearest isotropic tensor, whose parameters are $c^{\overline{\rm iso}}_{1111}=18.46\times10^6$ and $c^{\overline{\rm iso}}_{2323}=3.71\times10^6$\,, is obtained using expressions~(\ref{eq:ISO1111TI}) and (\ref{eq:ISO2323TI}).
The corresponding $P$-wave and $S$-wave speeds---which are the square roots of these parameters---are $v_P=4.30\,{\rm km\,s}^{-1}$ and $v_S=1.93\,{\rm km\,s}^{-1}$\,.

Following expressions~(\ref{eq:gamma}), (\ref{eq:delta}) and (\ref{eq:epsilon}), we obtain $\gamma=0.09$\,, $\delta=-0.04$ and $\epsilon=0.01$\,.
Since these values are close to zero, we conclude that---for the layer parameters in Table~\ref{table:BackusAverage}---the resulting Backus model is only weakly anisotropic.

\subsection{Strong inhomogeneity}
Let us consider a strongly inhomogeneous stack of alternating isotropic layers, whose elasticity parameters and velocities are given in Table~\ref{table:BackusAverage2}.

\begin{table}
\begin{center}
\begin{tabular}{|c||c|c|c|c|c|}
\hline
layer & $c_{1111}$ & $c_{2323}$ & $v_P$ & $v_S$ \\
\hline\hline
1 & 9 & 4 & 3 & 2 \\
2 & 49 & 16 & 7 & 4 \\
3 & 9 & 4 & 3 & 2 \\
4 & 49 & 16 & 7 & 4 \\
5 & 9 & 4 & 3 & 2 \\
6 & 49 & 16 & 7 & 4 \\
7 & 9 & 4 & 3 & 2 \\
8 & 49 & 16 & 7 & 4 \\
9 & 9 & 4 & 3 & 2 \\
10 & 49 & 16 & 7 & 4 \\
\hline
\end{tabular}
\end{center}
\caption{\small{Density-scaled elasticity parameters,~$\times 10^6\,{\rm m}^{2}\,{\rm s}^{-2}$\,, for a strongly inhomogeneous stack of alternating isotropic layers, and the corresponding $P$-wave and $S$-wave velocities,~${\rm km\,s}^{-1}$}}
\label{table:BackusAverage2}
\end{table}

\begin{figure}
\begin{center}
\includegraphics[scale=0.55]{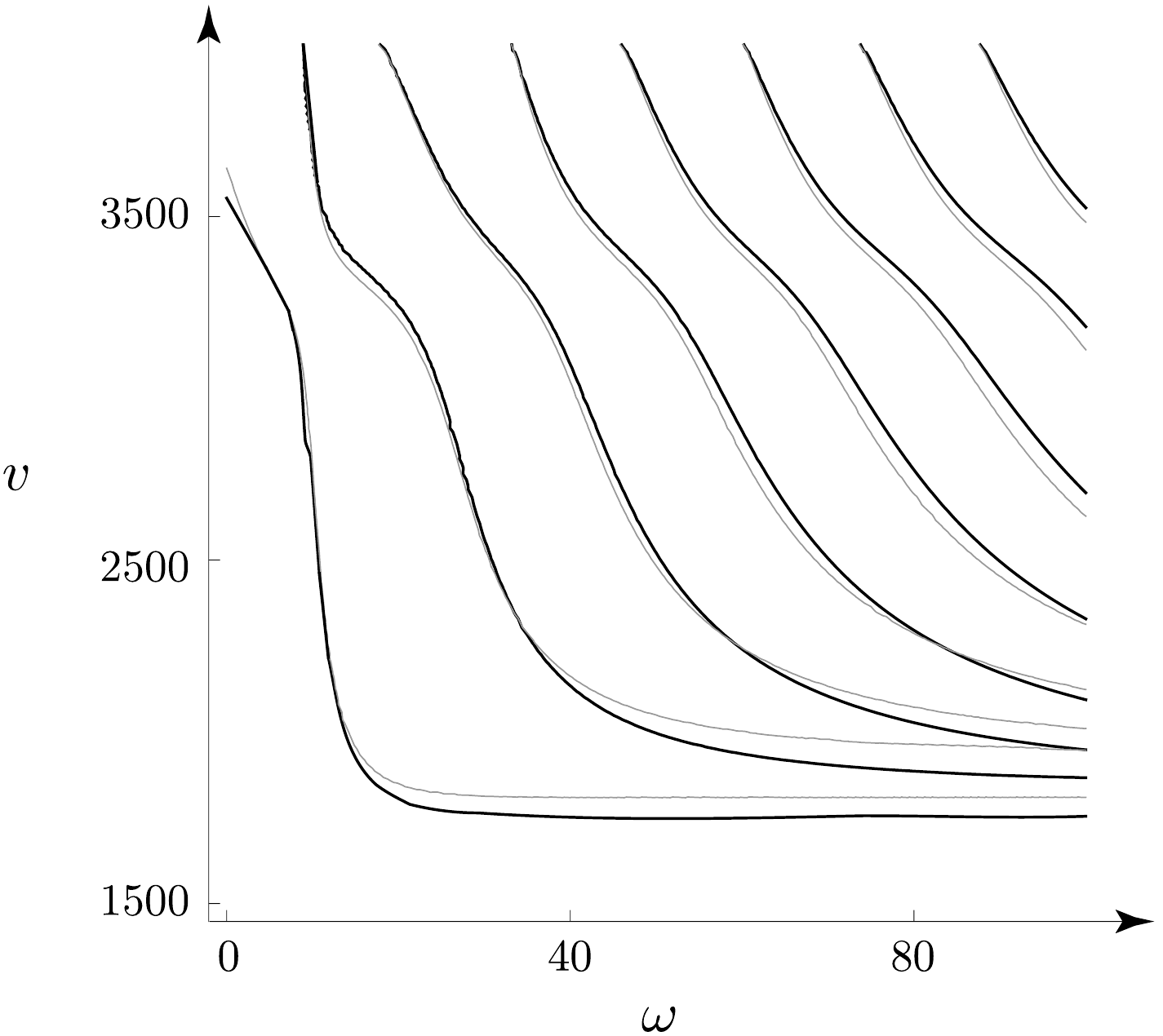}
\includegraphics[scale=0.55]{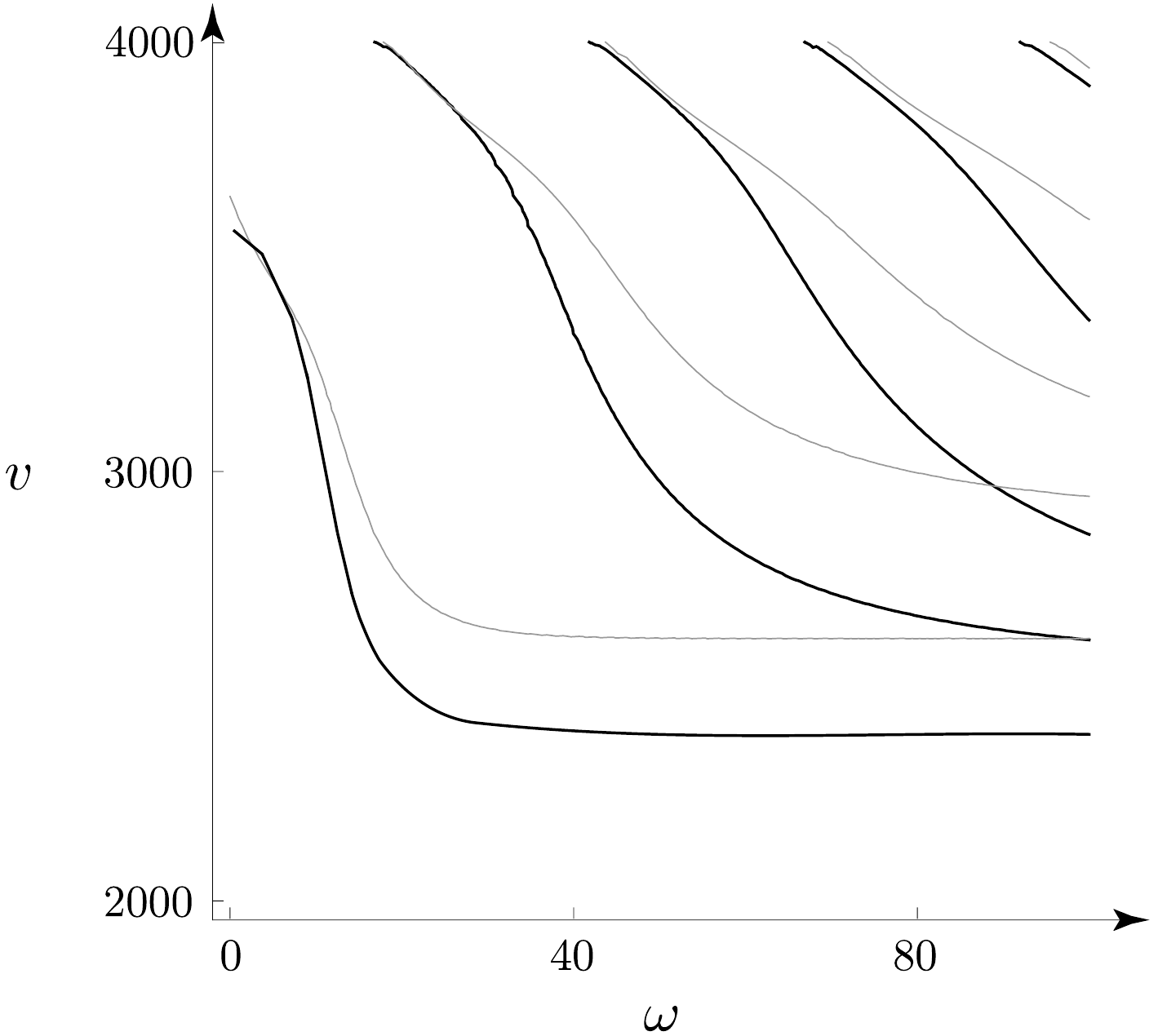}
\end{center}
\caption{\small{quasi-Rayleigh wave dispersion curves for the Backus medium and the Voigt medium, shown as black and grey lines, respectively.
}}
\label{fig:TI-vs-iso-R}
\end{figure}

\begin{figure}
\begin{center}
\includegraphics[scale=0.55]{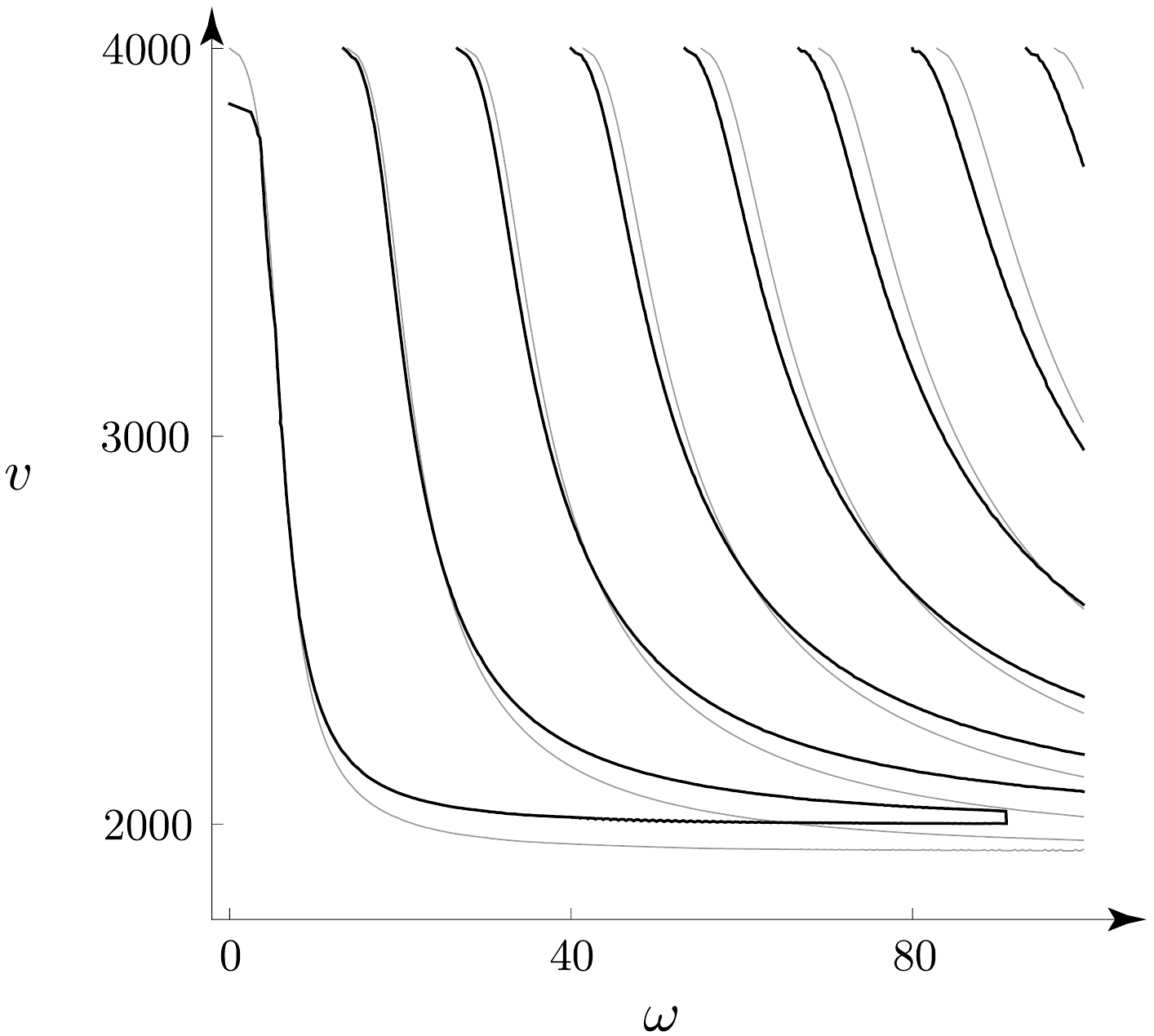}
\includegraphics[scale=0.55]{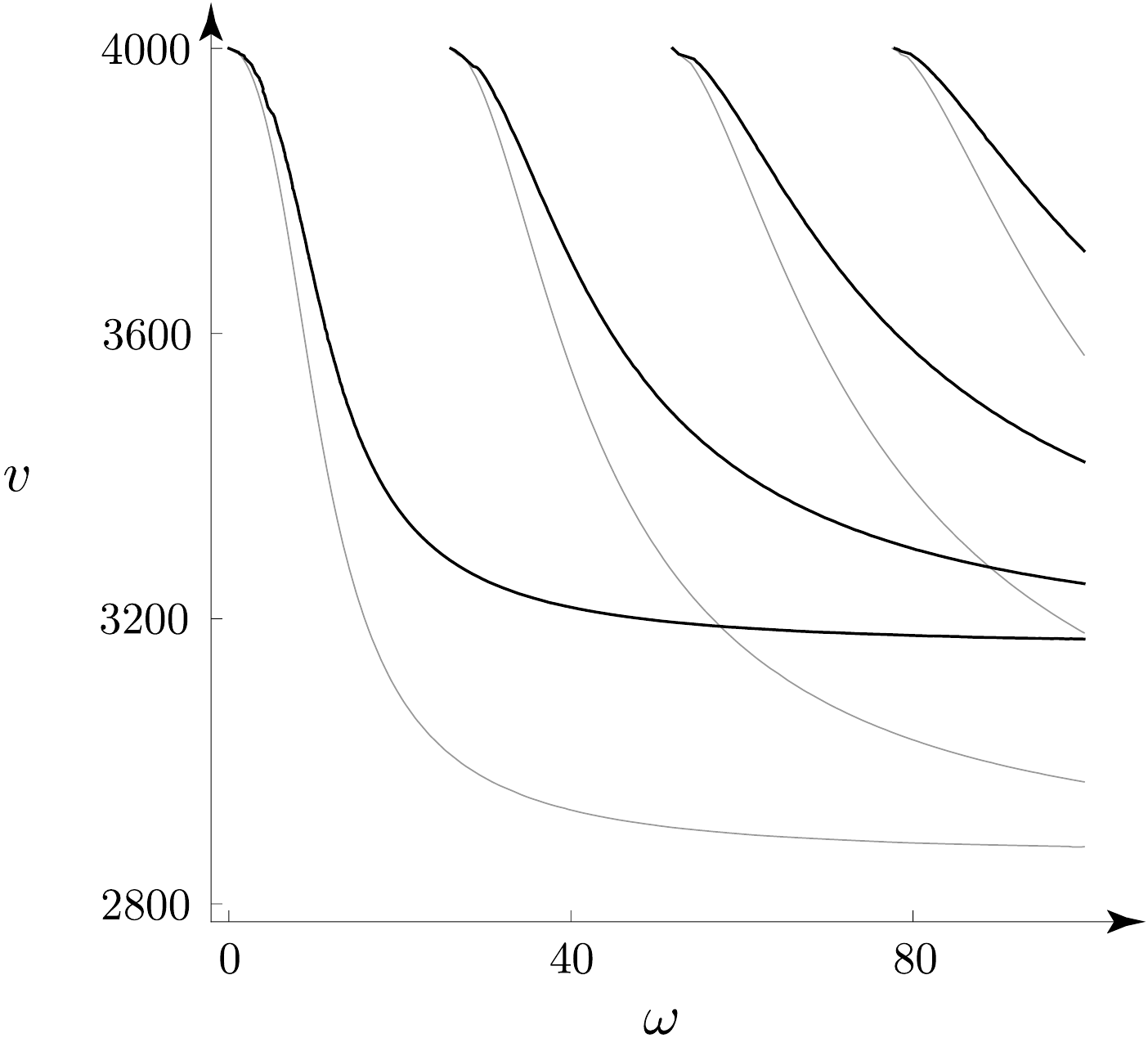}
\end{center}
\caption{\small{Love wave dispersion curves for the Backus medium and the Voigt medium, shown as black and grey lines, respectively}}
\label{fig:TI-vs-iso-L}
\end{figure}

Following expressions~(\ref{eq:Tue1})--(\ref{eq:Tue2}), we obtain
$c^{\overline{\rm TI}}_{1111}=26.79$\,,
$c^{\overline{\rm TI}}_{1133}=3.48$\,,
$c^{\overline{\rm TI}}_{1212}=10.00$\,,
$c^{\overline{\rm TI}}_{2323}=6.40$ and
$c^{\overline{\rm TI}}_{3333}=15.21$\,;
these values are to be multiplied by $10^6$\,, and their units are ${\rm m}^{2}\,{\rm s}^{-2}$\,.
Following expressions~(\ref{eq:ISO1111TI}) and (\ref{eq:ISO2323TI}), we obtain $c^{\overline{\rm iso}}_{1111}=21.67\times10^6$ and $c^{\overline{\rm iso}}_{2323}=8.23\times10^6$\,, and the corresponding $P$-wave and $S$-wave speeds are $v_P=4.66\,{\rm km\,s}^{-1}$ and $v_S=2.87\,{\rm km\,s}^{-1}$\,.

Following expressions~(\ref{eq:gamma}), (\ref{eq:delta}) and (\ref{eq:epsilon}), we obtain $\gamma=0.28$\,, $\delta=0.08$ and $\epsilon=0.38$\,.
Since these values are not close to zero, we conclude that---for the layer parameters in Table~\ref{table:BackusAverage2}---the resulting Backus model is strongly anisotropic.
\subsection{Discussion}
\begin{figure}
\begin{center}
\includegraphics[scale=0.55]{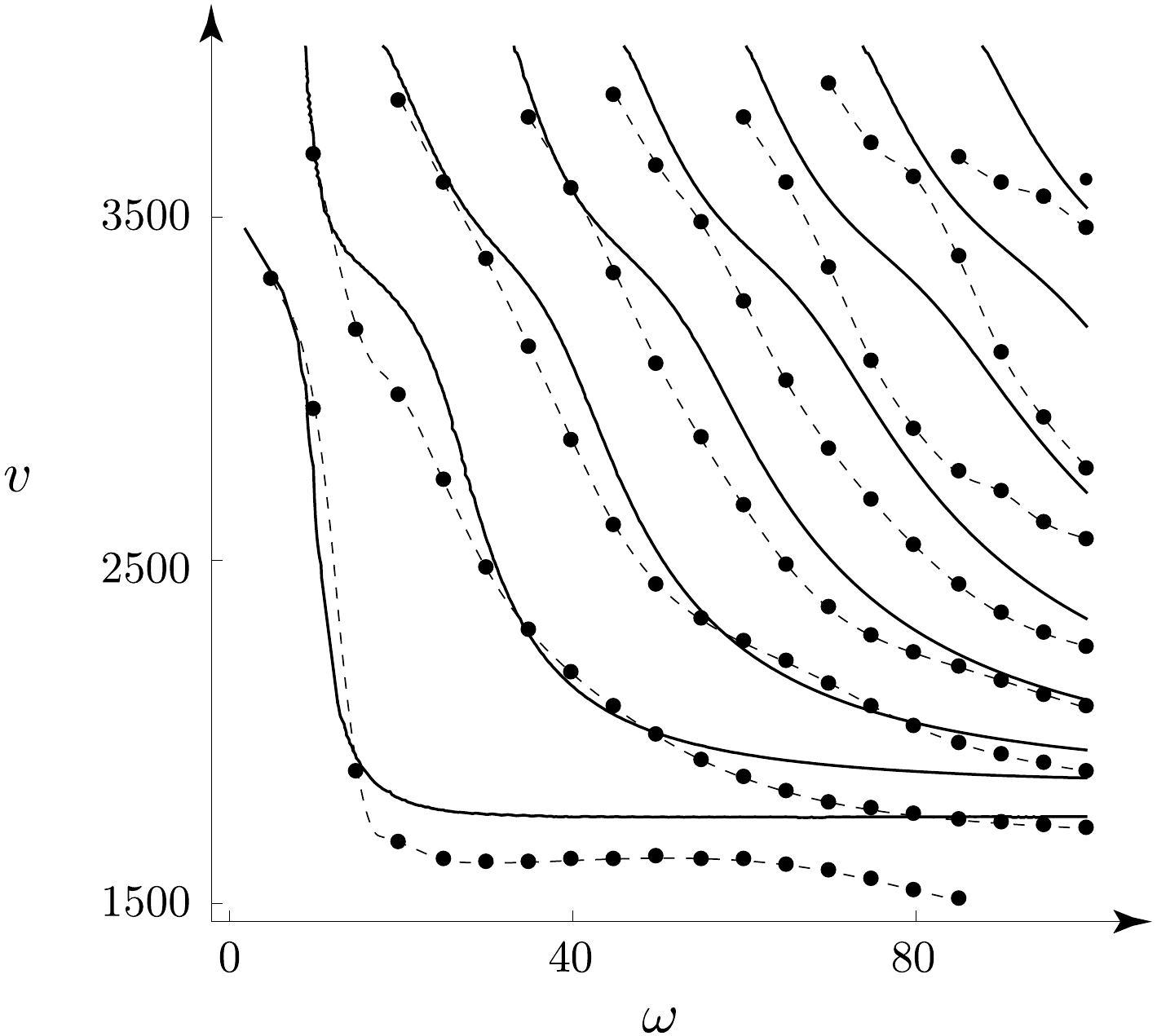}\includegraphics[scale=0.55]{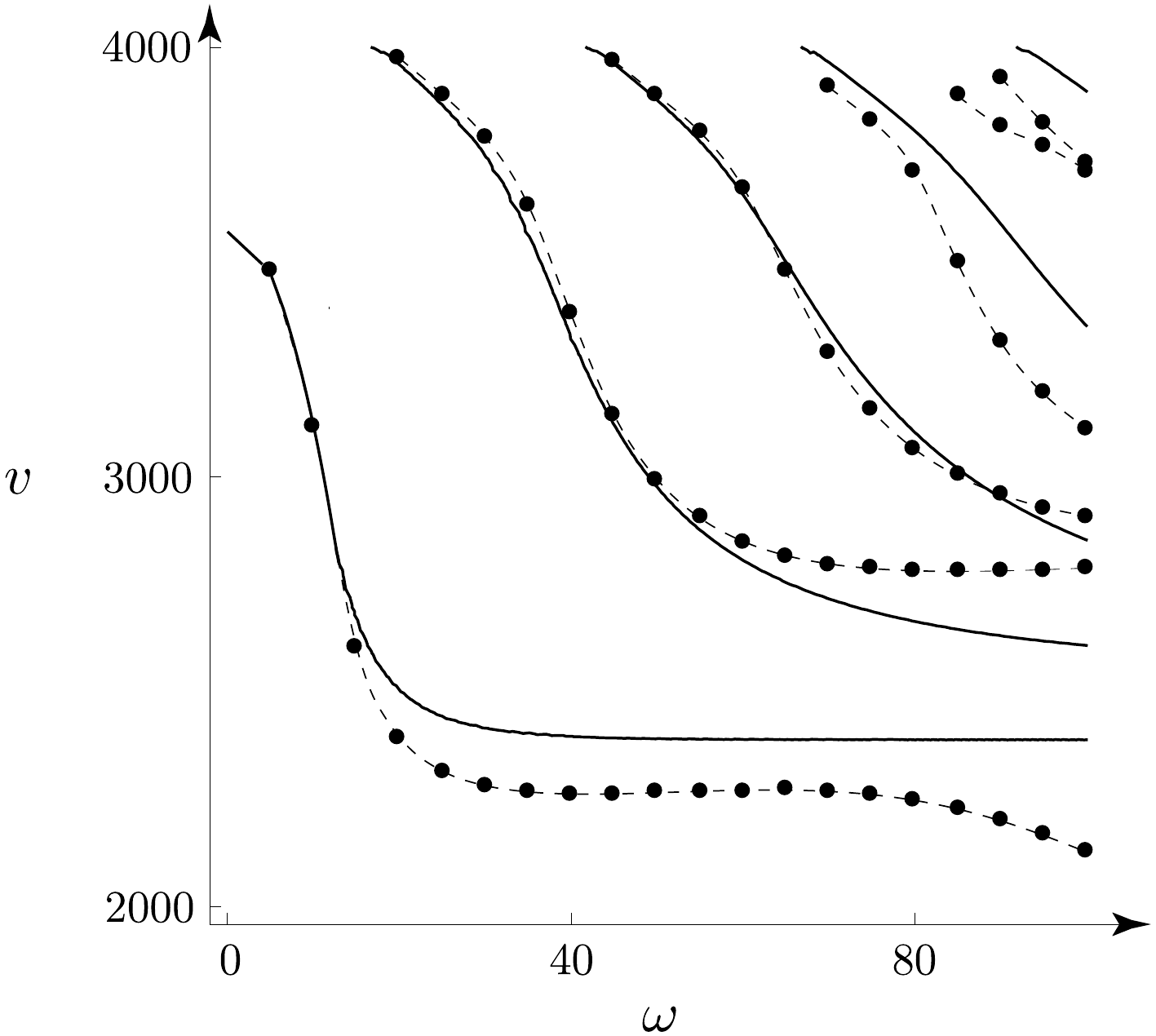}
\end{center}
\caption{\small{quasi-Rayleigh wave dispersion curves for the Backus medium and the delta-matrix solution, shown as black lines and points, respectively, for layers that are fifty-metres thick}}
\label{fig:CTI500-R}
\end{figure}

\begin{figure}
\begin{center}
\includegraphics[scale=0.55]{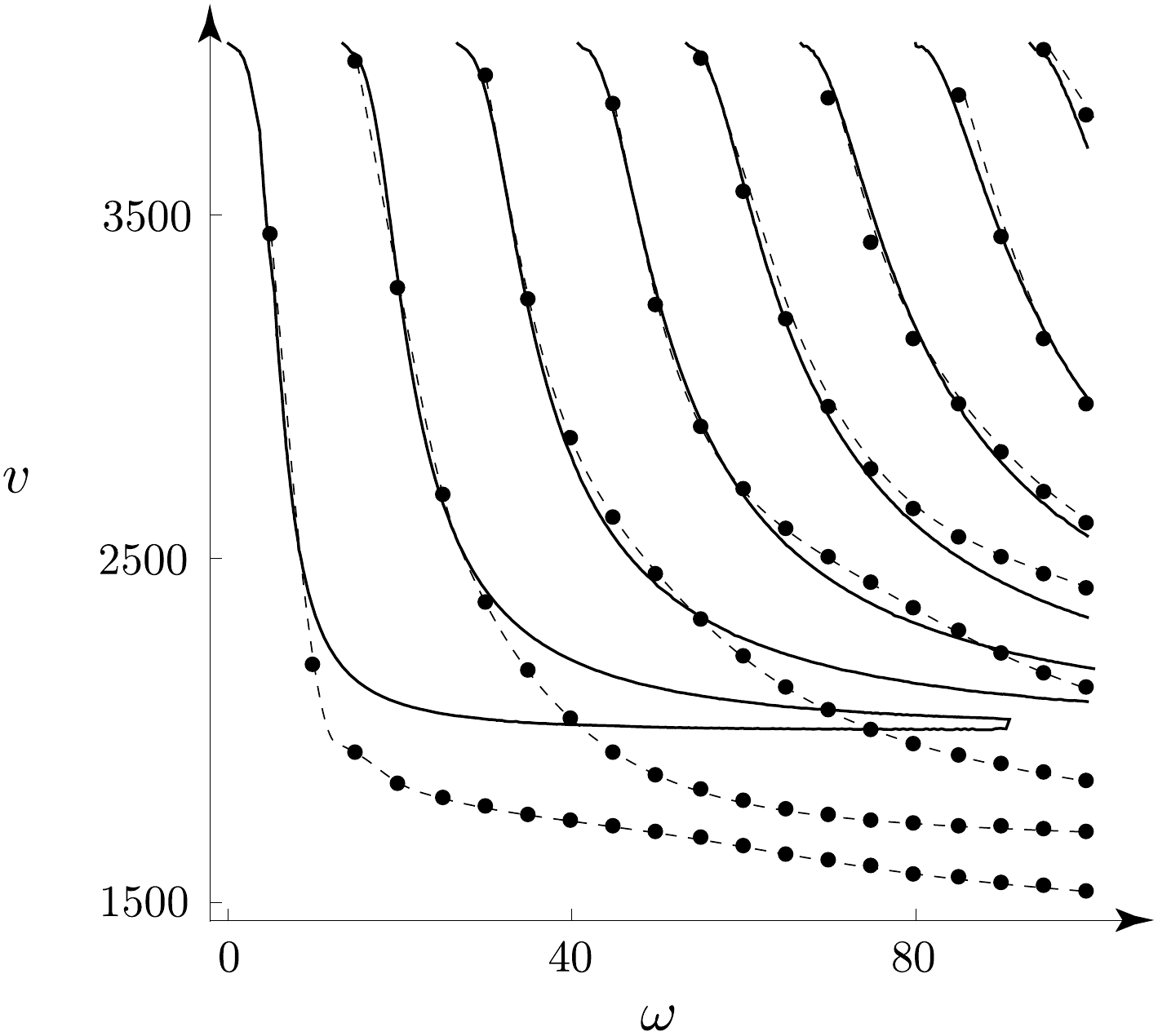}
\includegraphics[scale=0.55]{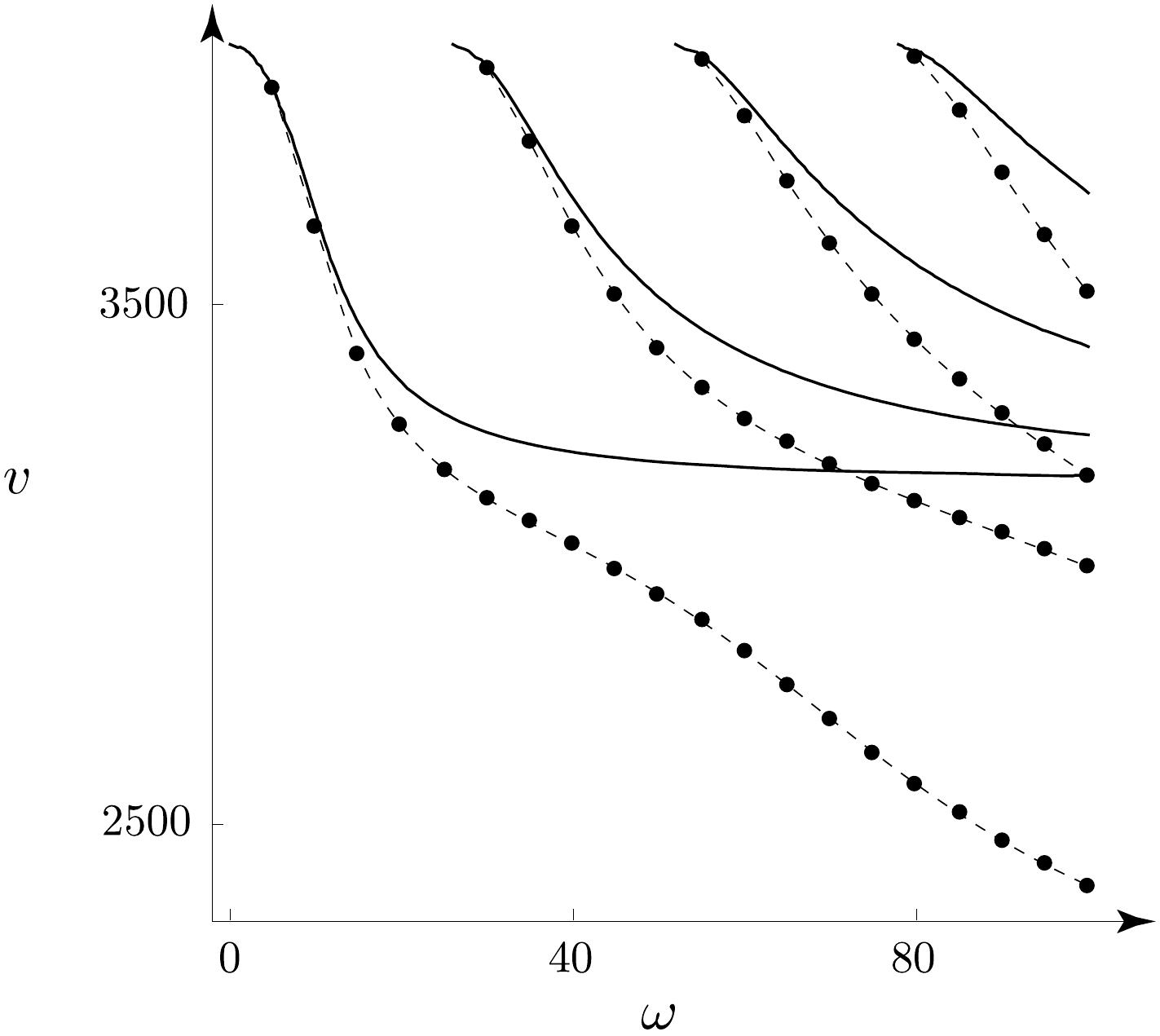}
\end{center}
\caption{\small{Love wave dispersion curves for the Backus medium and the delta-matrix solution, shown as black lines and points, respectively, for layers that are fifty-metres thick
}}
\label{fig:CTI500-L}
\end{figure}

\begin{figure}
\begin{center}
\includegraphics[scale=0.55]{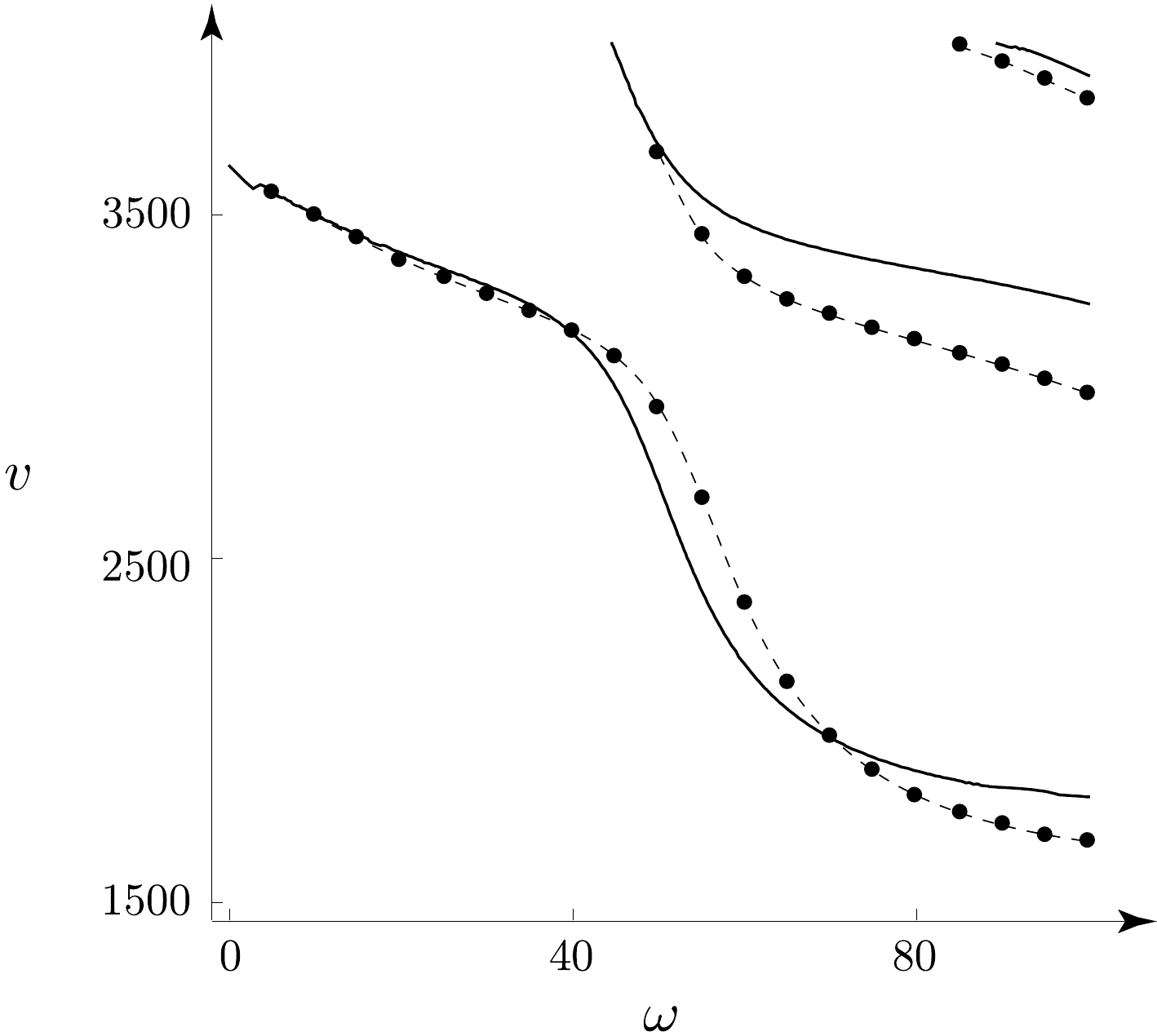}\includegraphics[scale=0.55]{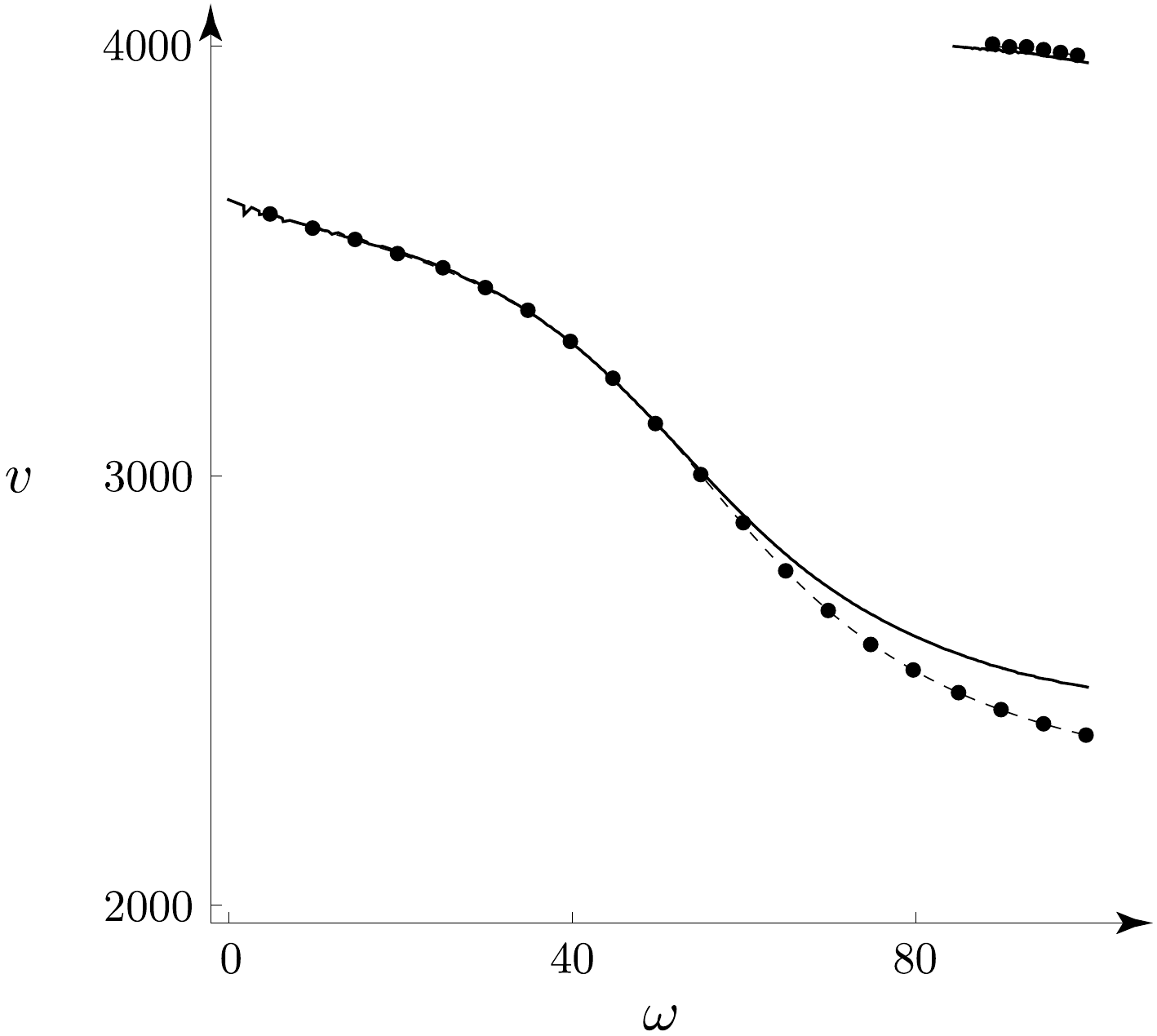}
\end{center}
\caption{\small{quasi-Rayleigh wave dispersion curves for the Backus medium and the delta-matrix solution, shown as black lines and points, respectively, for layers that are ten-metres thick
}}
\label{fig:CTI100-R}
\end{figure}

\begin{figure}
\begin{center}
\includegraphics[scale=0.55]{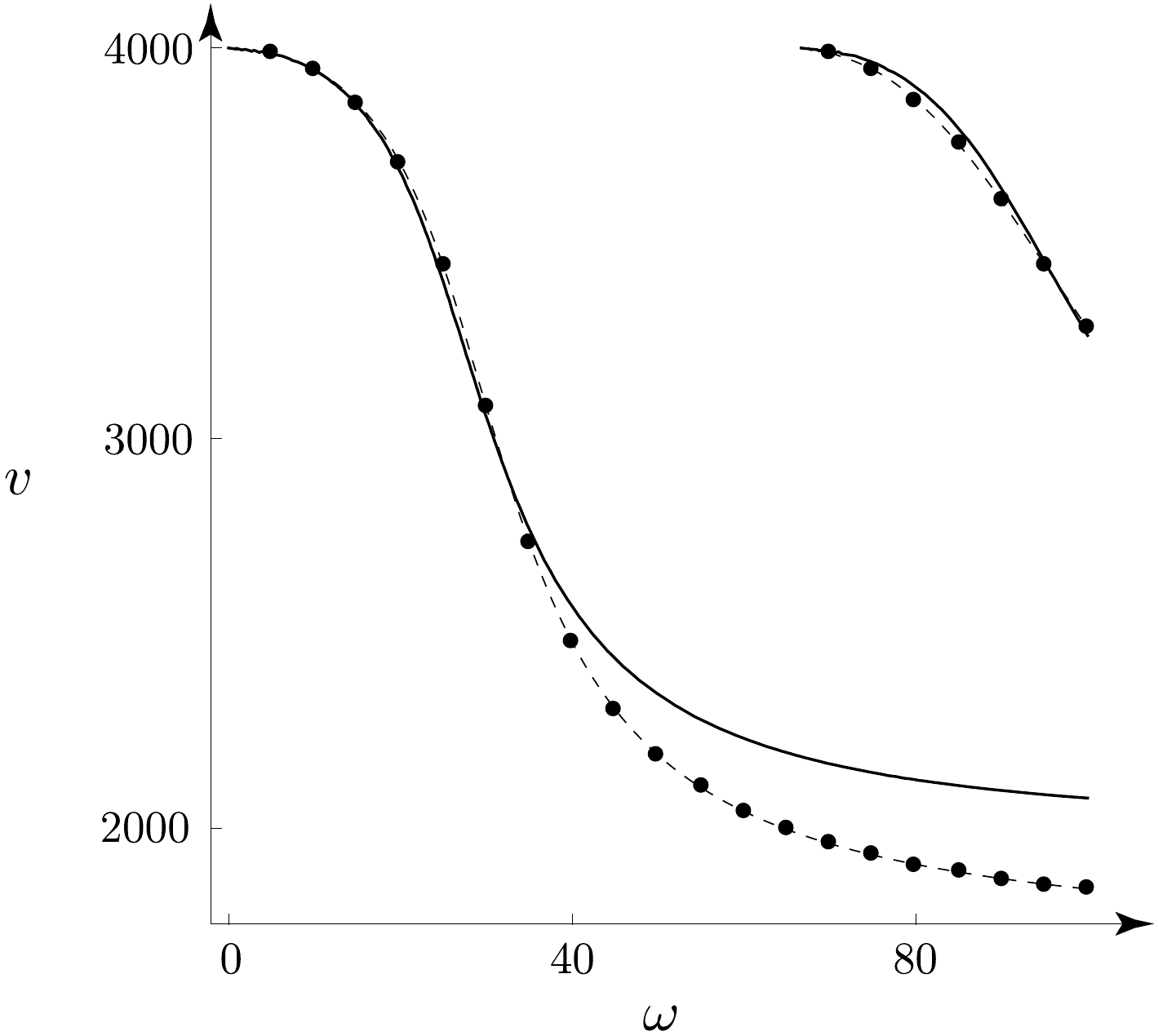}
\includegraphics[scale=0.55]{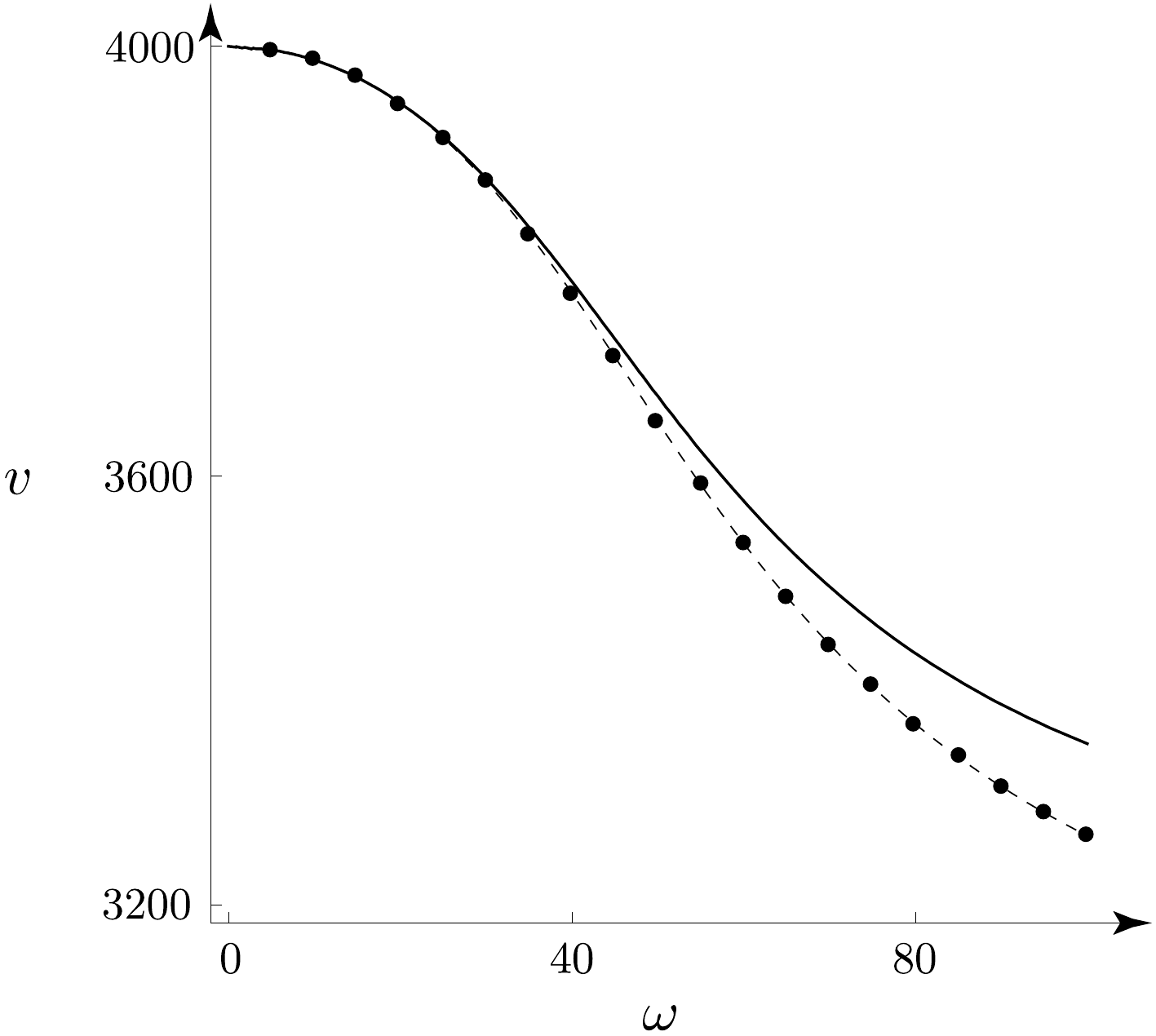}
\end{center}
\caption{\small{Love wave dispersion curves for the Backus medium and the delta-matrix solution, shown as black lines and points, respectively, for layers that are ten-metres thick}}
\label{fig:CTI100-L}
\end{figure}

\begin{figure}
\begin{center}
\includegraphics[scale=0.55]{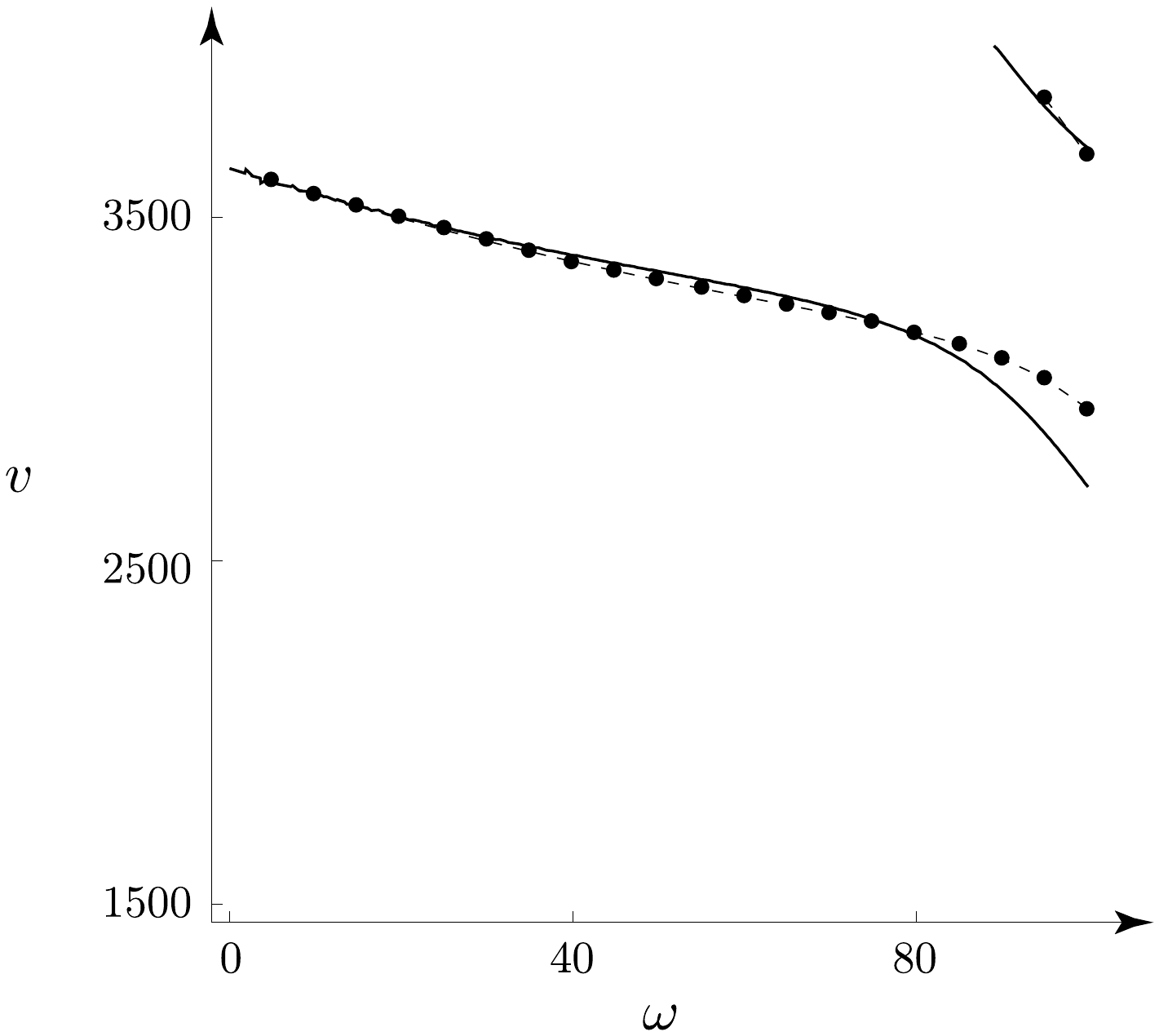}
\includegraphics[scale=0.55]{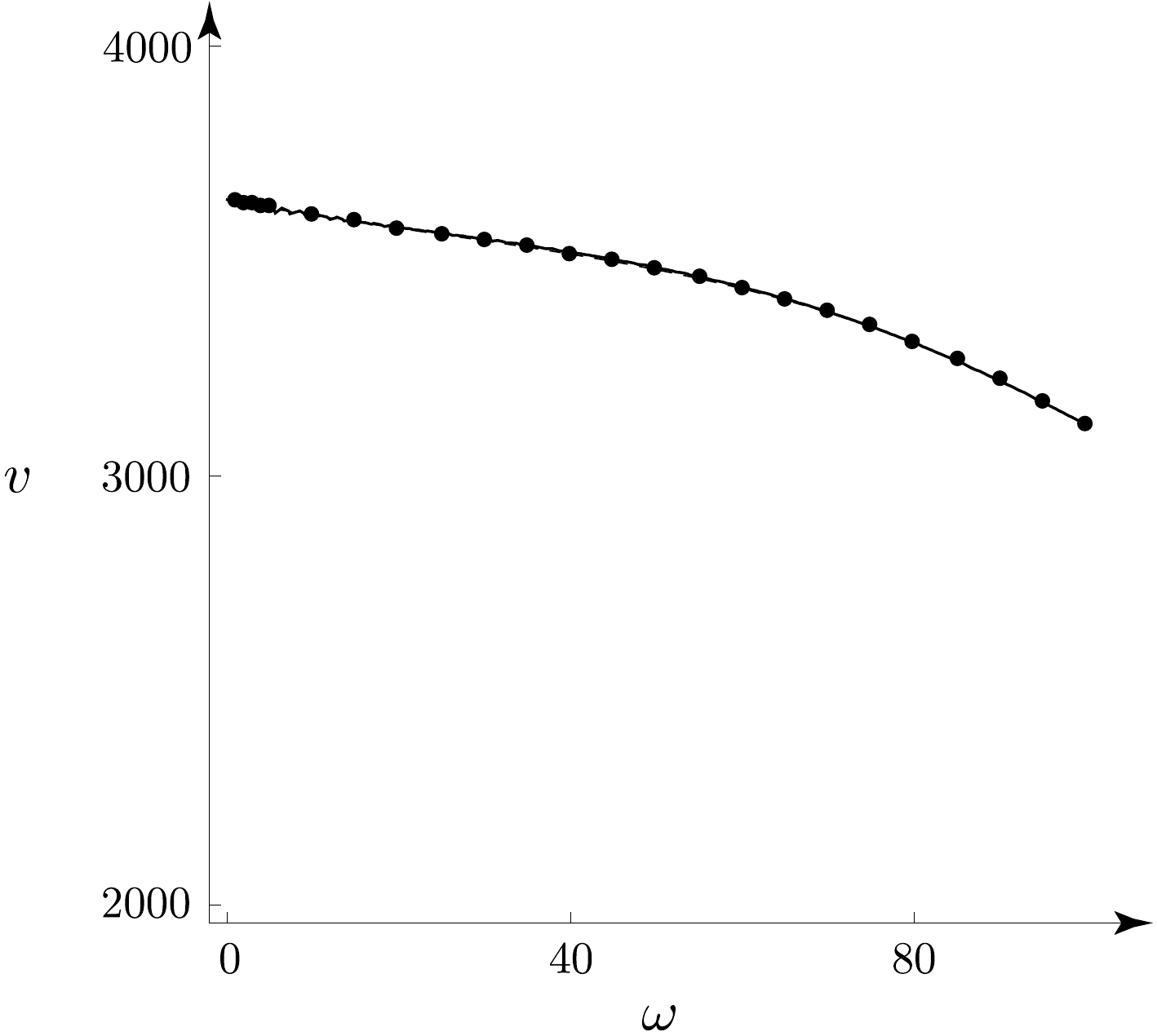}
\end{center}
\caption{\small{quasi-Rayleigh wave dispersion curves for the Backus medium and the delta-matrix solution, shown as black lines and points, respectively, for layers that are five-metres thick
}}
\label{fig:CTI50-R}
\end{figure}

\begin{figure}
\begin{center}
\includegraphics[scale=0.55]{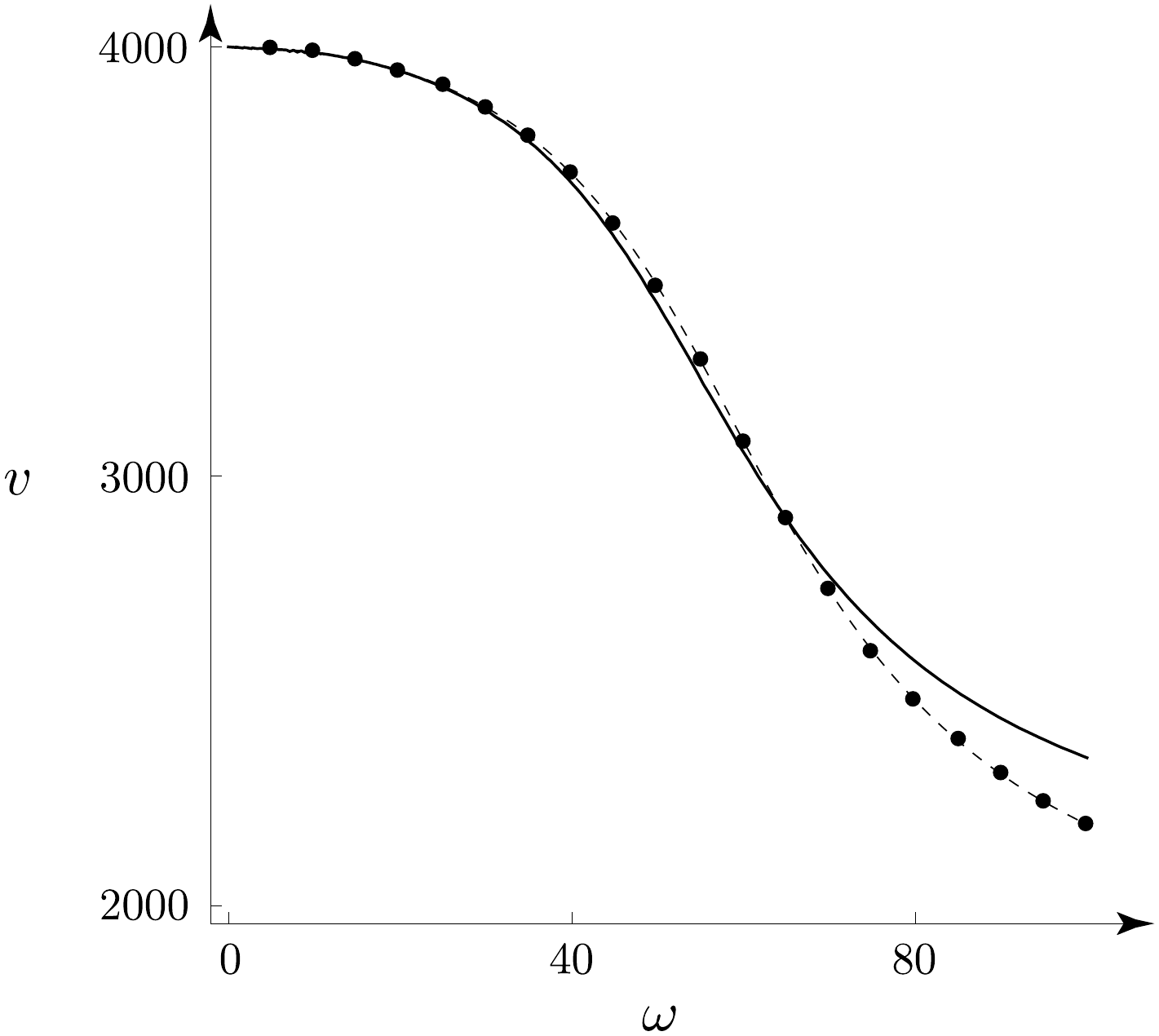}
\includegraphics[scale=0.55]{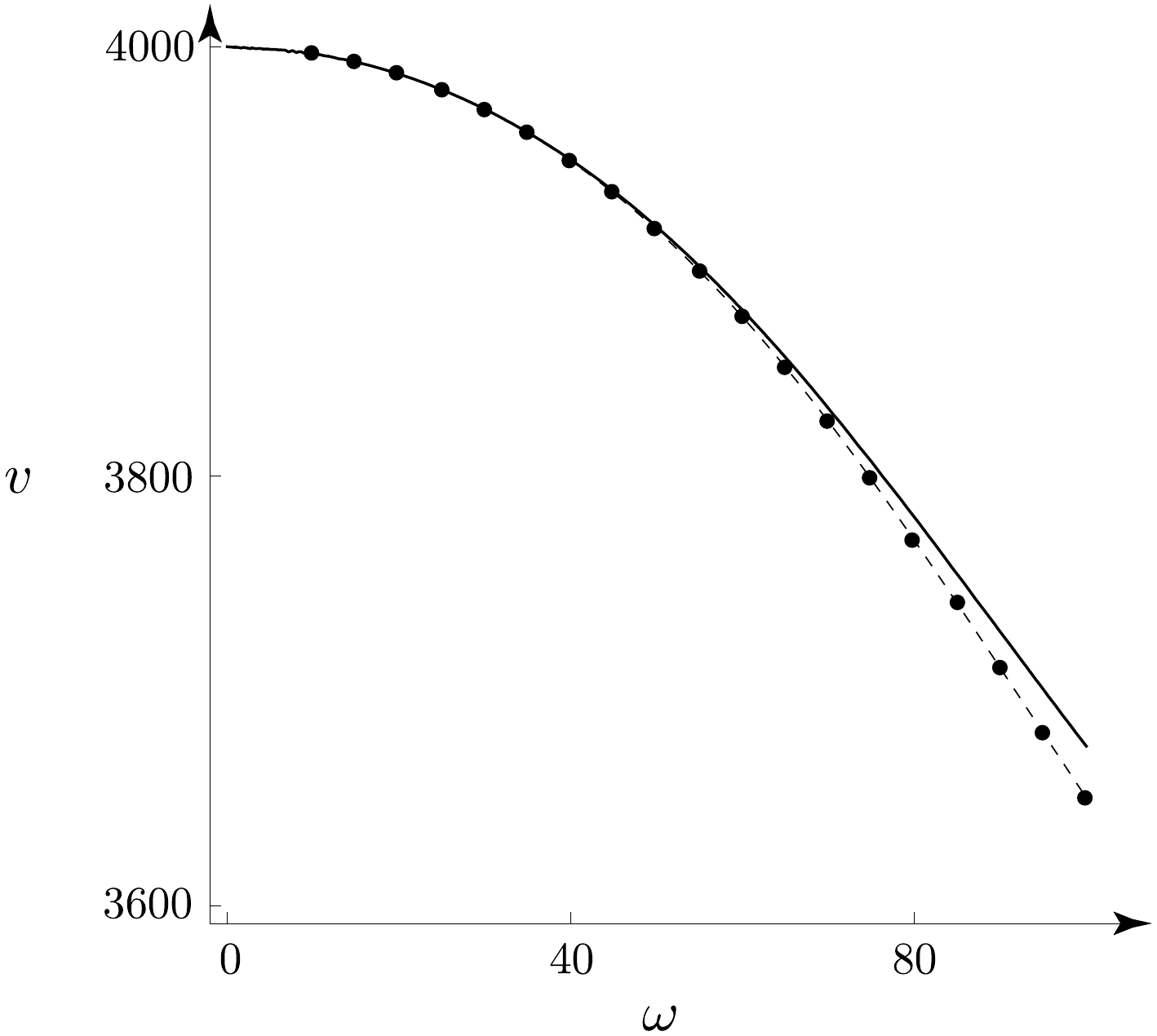}
\end{center}
\caption{\small{Love wave dispersion curves for the Backus medium and the delta-matrix solution, shown as black lines and points, respectively, for layers that are five-metres thick
}}
\label{fig:CTI50-L}
\end{figure}

Figures~\ref{fig:TI-vs-iso-R}--\ref{fig:CTI50-L} illustrate the dispersion curves for Love and quasi-Rayleigh waves for Backus media that are averages of isotropic layers, for Voigt media that are averages of these Backus media, and for delta-matrix solutions for the stack of these layers.
In each figure, the left-hand plot exhibits a weakly anisotropic Backus medium and the right-hand plots exhibits a strongly anisotropic Backus medium; they are compared with either a Voigt medium or a delta-matrix solution.

In each plot, the lowest curve corresponds to the fundamental mode of a guided wave, the next curve to its first mode, and so on.
For the fundamental mode, the speed values correspond to all frequencies,~$(0,\infty)$\,; higher modes have cutoff frequencies, whose values are greater than zero; the speeds of these modes correspond to $(\omega_0,\infty)$\,, where $\omega_0>0$\,.
At the zero-frequency limit, the dispersion relation is affected only by properties of the halfspace; at the high-frequency limit, it is affected only by properties of the overlying medium; the intermediate region is affected by properties of the entire model.
For the quasi-Rayleigh wave, the lower limit is the value of the Rayleigh-wave speed in the halfspace.
For the Love wave, it is the $S$-wave speed in the halfspace.
These issues are discussed by Dalton et al.~\cite{dalton} and Ud\'{\i}as~\cite[p.~201]{Udias1999}.
However, we note that the highest frequency available in a seismic signal might not allow us to reach the required upper limit.


For both weakly and strongly anisotropic media, the mass density of the halfspace, which is isotropic, is
$\rho^d=2600~{\rm kg}/{\rm m}^3$\,, and its  elasticity parameters are $c_{1111}^d= 10.99\times 10^{10}~{\rm N}/{\rm m}^2$ and  $c_{2323}^d=4.16\times 10^{10}~{\rm N}/{\rm m}^2$\,.
The mass density of the layers---and, hence, of the resulting medium---is $\rho^u=2200~{\rm kg}/{\rm m}^3$\,.
In Figures~\ref{fig:TI-vs-iso-R}--\ref{fig:CTI500-L}, the medium overlying the isotropic halfspace is five-hundred-metres thick; in Figures~\ref{fig:CTI100-R} and~\ref{fig:CTI100-L}, it is one-hundred-metres thick; in Figures~\ref{fig:CTI50-R} and~\ref{fig:CTI50-L}, it is fifty-metres thick.
Since, in each case, the Backus~\cite{backus} average is taken over ten layers, the resulting Backus-medium parameters are not affected by the value of the total thickness, as illustrated by expression~(\ref{eq:weight}).
This value, however, affects the dispersion relations, as exemplified---for the isotropic case---by the presence of $Z$ in expressions~(2) and~(17) of Dalton et al.~\cite{dalton} . 

Examining Figures~\ref{fig:TI-vs-iso-R} and \ref{fig:TI-vs-iso-L}, we observe the effect of layer inhomogeneity on the dispersion curves of the quasi-Rayleigh and Love waves.
As illustrated by the similarity of curves for the Backus and Voigt media, in the left-hand plots, the effect is negligible under weak inhomogeneity, which---for the Backus~\cite{backus} average---becomes weak anisotropy.
As indicated by the discrepancy between these curves, in the right-hand plots, the effect, which is negligible near the cutoff frequency, becomes pronounced with the increase of frequency. 
The match is good at lower frequencies, since the dispersion relation is dominated by properties of the halfspace.

Examining Figures~\ref{fig:CTI500-R}, \ref{fig:CTI100-R}, \ref{fig:CTI50-R}, we see that there is a good match between the Backus media---both weakly and strongly anisotropic---and the delta-matrix solutions, but only for thinner layers or lower frequencies.   
Also, there is a good match in the second, third, and fourth modes for frequencies near the cutoff frequency for the thick layers in Figure~\ref{fig:CTI500-R}, where the speed approaches the $S$-wave speed in the halfspace.
A similar pattern appears in Figures~\ref{fig:CTI500-L}, \ref{fig:CTI100-L}, \ref{fig:CTI50-L}.

Since both quasi-Rayleigh and Love waves exist in the model consisting of the same parameters, we can examine their dispersion curves that correspond to the same model.
These are shown in Figures~\ref{fig:CTI500-R} and \ref{fig:CTI500-L}, Figures~\ref{fig:CTI100-R} and \ref{fig:CTI100-L}, Figures~\ref{fig:CTI50-R} and~\ref{fig:CTI50-L}; each respective pair corresponds to a model of fifty-, ten- and five-metre-thick layers.
Examining Figures~\ref{fig:CTI500-R} and \ref{fig:CTI500-L}, for instance, we see that---for both weak and strong inhomogeneity---the match between the results of the Backus~\cite{backus} average and the propagator matrix appears to be better for all modes of the Love wave, at lower frequencies, but, in the case of the quasi-Rayleigh wave, extends to higher frequencies.
\section{Transversely isotropic layers}
For a Backus medium transversely isotropic layer, the dispersion relations for quasi-Rayleigh and Love waves are given by setting to zero the determinants of the matrices in equations~(29) and~(30), respectively, of Khojasteh et al.~\cite{khojasteh}.
These relations can also be derived by setting to zero the thickness of the liquid layer in equations~(22) and~(23) of Bagheri et al.~\cite{bagheri}.
In a manner analogous to expressions~(2) and~(17) of Dalton et al.~\cite{dalton} the properties of the layer consist of its thickness, mass density, and five elasticity parameters of a transversely isotropic continuum, $c_{1111}$\,, $c_{1122}$\,, $c_{1133}$\,, $c_{1133}$ and $c_{2323}$\,.
The halfspace could be isotropic, in which case its properties would be the same as in expressions~(2) and~(17) of Dalton et al.~\cite{dalton}; it could be also transversely isotropic.

Again, these relations are coded in Mathematica\textsuperscript{\textregistered}, but the dispersion curves are plotted as zero contours of the sum of the real and imaginary parts of the respective determinants.

For a stack of transversely isotropic layers overlying a transversely isotropic halfspace, the dispersion relation for quasi-Rayleigh waves is based on the reduced-delta-matrix solution of Ikeda and Matsuoka~\cite{ikedamatsuoka}, and the dispersion relation for Love waves is based on the  delta-matrix solution reviewed in Buchen and Ben-Hador~\cite{buchen} but with pseudorigidity and pseudothickness defined as in Anderson~\cite{anderson62}.
These relations are coded in Python\textsuperscript{\textregistered}.
The algorithm is similar to the delta-matrix solution, except with expressions for transversely isotropic continua, which contain five elasticity parameters, as derived by Anderson~\cite{anderson62} and Ikeda and Matsuoka~\cite{ikedamatsuoka}.

\subsection{Alternating layers on isotropic halfspace}
Let us consider a stack of alternating transversely isotropic layers,
given in Table~\ref{table:altTI}, where the parameters of the odd-numbered
layers are from tensor $C_a^{\rm TI}$ of Danek et al.~\cite{DNS} and the
parameters of the even-numbered layers are twice those of the tensor
$C_{bb}^{\rm TI}$ of Danek et al.~\cite{DNS}.   
We chose these parameters to illustrate varying levels of anisotropy, quantified by parameters~(\ref{eq:gamma}), (\ref{eq:delta}) and (\ref{eq:epsilon}).
We set the mass density of the layers to $2200~{\rm kg}/{\rm m}^3$\,, and use the same halfspace parameters as in Section~\ref{sec:IsoLay}.

\begin{table}
\begin{center}
\begin{tabular}{|c||c|c|c|c|c|}
\hline
layer & $c_{1111}$ & $c_{1133}$ & $c_{3333}$ & $c_{2323}$ & $c_{1212}$ \\
\hline\hline
1 & 8.06 & 2.46 & 7.08 & 1.86 & 2.35\\
2 & 13.73 & 5.75 &16.77& 5.55 & 3.56\\
3 & 8.06  & 2.46 & 7.08 & 1.86 & 2.35\\
4 & 13.73 & 5.75 &16.77& 5.55 & 3.56\\
5 & 8.06  & 2.46 & 7.08 & 1.86 & 2.35\\
6 & 13.73 & 5.75 &16.77& 5.55 & 3.56\\
7 & 8.06  & 2.46 & 7.08 & 1.86 & 2.35\\
8 & 13.73 & 5.75 &16.77& 5.55 & 3.56\\
9 & 8.06  & 2.46 & 7.08 & 1.86 & 2.35\\
10 & 13.73 & 5.75 &16.77& 5.55 & 3.56\\
\hline
\end{tabular}
\end{center}
\caption{\small{Density-scaled elasticity parameters, whose units are $10^6\,{\rm m}^{2}{\rm s}^{-2}$\,, for a stack of alternating transversely isotropic layers}}
\label{table:altTI}
\end{table}

Following expressions~(\ref{eq:c1111})--(\ref{eq:c3333}), we obtain
$c^{\overline{\rm TI}}_{1111}=10.67$\,,
$c^{\overline{\rm TI}}_{1133}=3.44$\,,
$c^{\overline{\rm TI}}_{1212}=2.95$\,,
$c^{\overline{\rm TI}}_{2323}=2.79$ and
$c^{\overline{\rm TI}}_{3333}=9.96$\,;
these values are multiplied by $10^6$\,, and their units are ${\rm m}^{2}{\rm s}^{-2}$\,.
Following expressions~(\ref{eq:ISO1111TI}) and (\ref{eq:ISO2323TI}), we have $c^{\overline{\rm iso}}_{1111}=10.09\times10^6$ and $c^{\overline{\rm iso}}_{2323}=3.02\times10^6$\,, respectively, which correspond to $v_P=3.27\,{\rm km\,s}^{-1}$ and $v_S=1.74\,{\rm km\,s}^{-1}$\,.
According to expressions~(\ref{eq:gamma}), (\ref{eq:delta}) and (\ref{eq:epsilon}), $\gamma=0.03$\,, $\delta=-0.09$ and $\epsilon=0.04$\,, respectively.
Since these values are close to zero, we conclude that---for the layer parameters in Table~\ref{table:altTI}---the resulting Backus model is only weakly anisotropic.

\begin{figure}
\begin{center}
\includegraphics[scale=0.55]{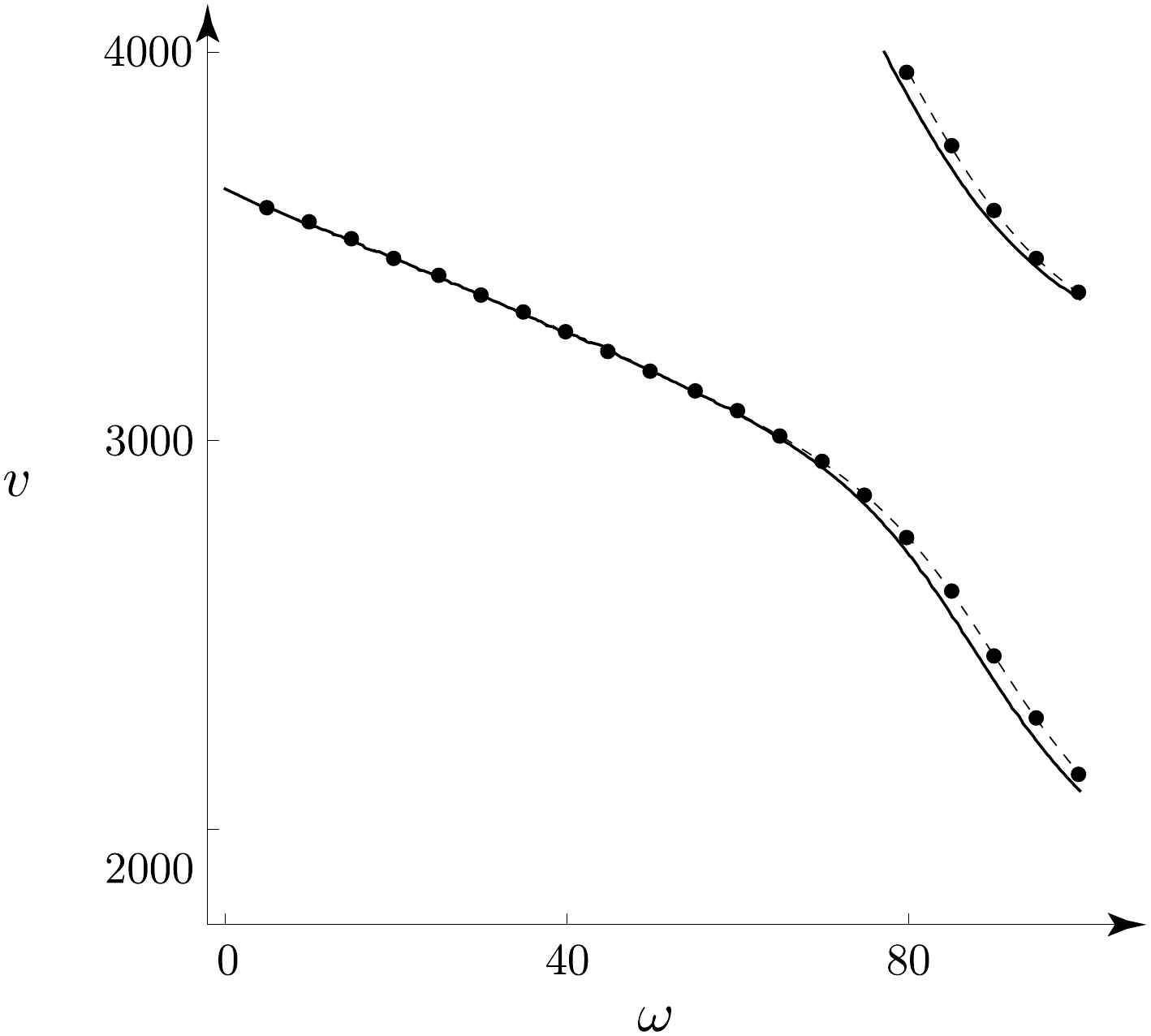}
\includegraphics[scale=0.55]{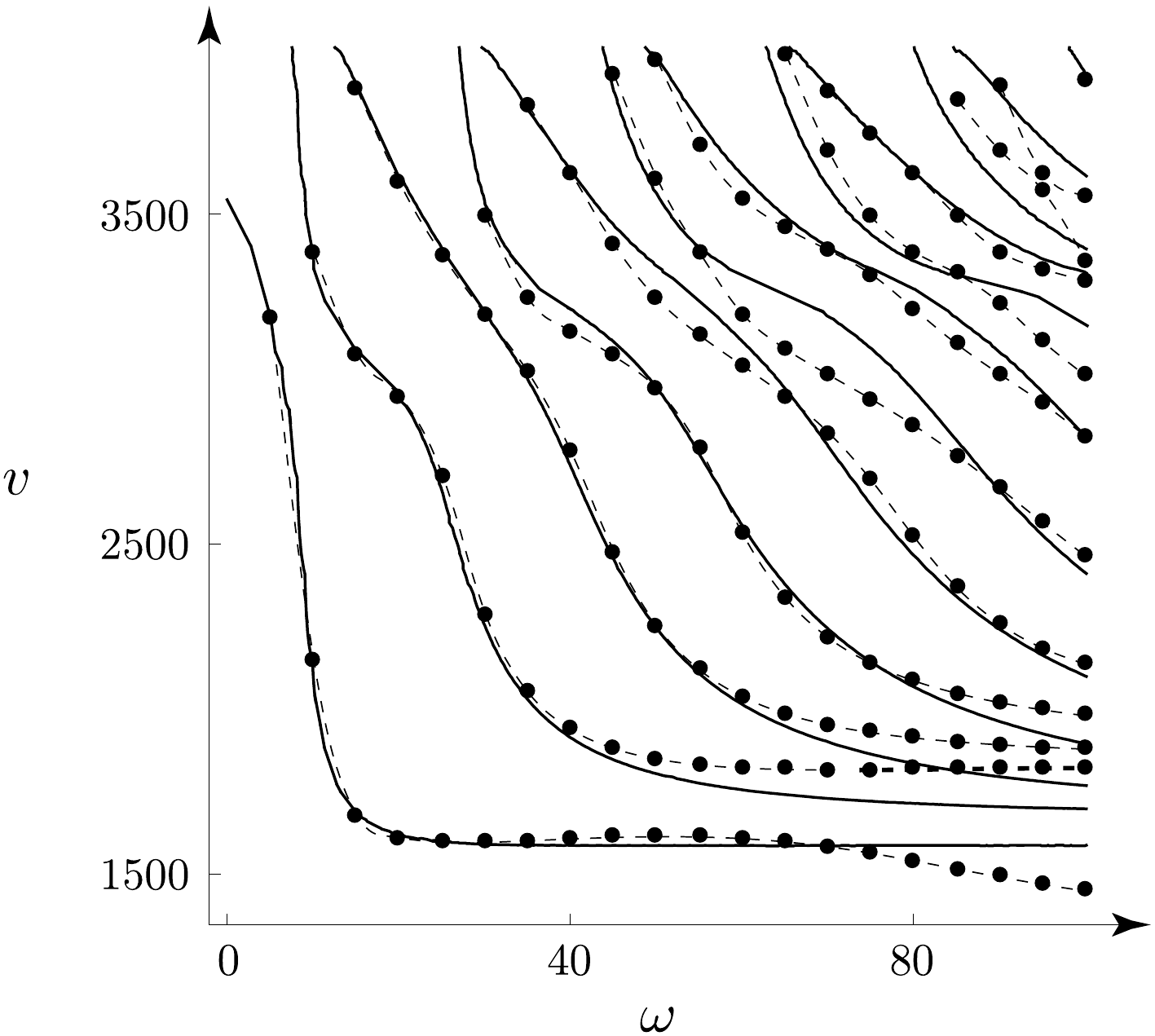}
\end{center}
\caption{\small{quasi-Rayleigh wave dispersion curves for
the Backus medium and reduced-delta-matrix solution, shown as black lines and points, respectively}}
\label{fig:qRTI}
\end{figure}
\begin{figure}
\begin{center}
\includegraphics[scale=0.55]{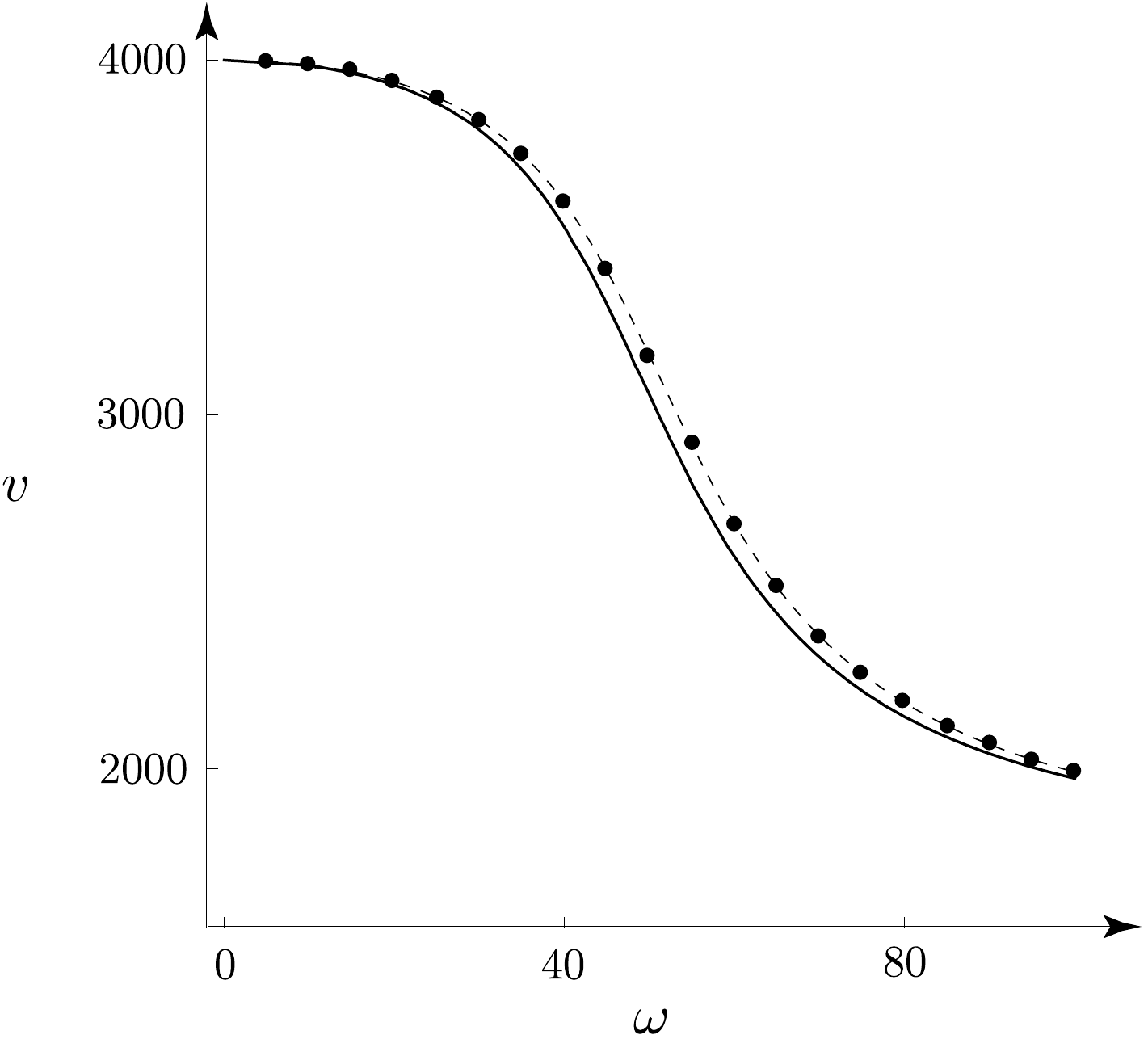}
\includegraphics[scale=0.55]{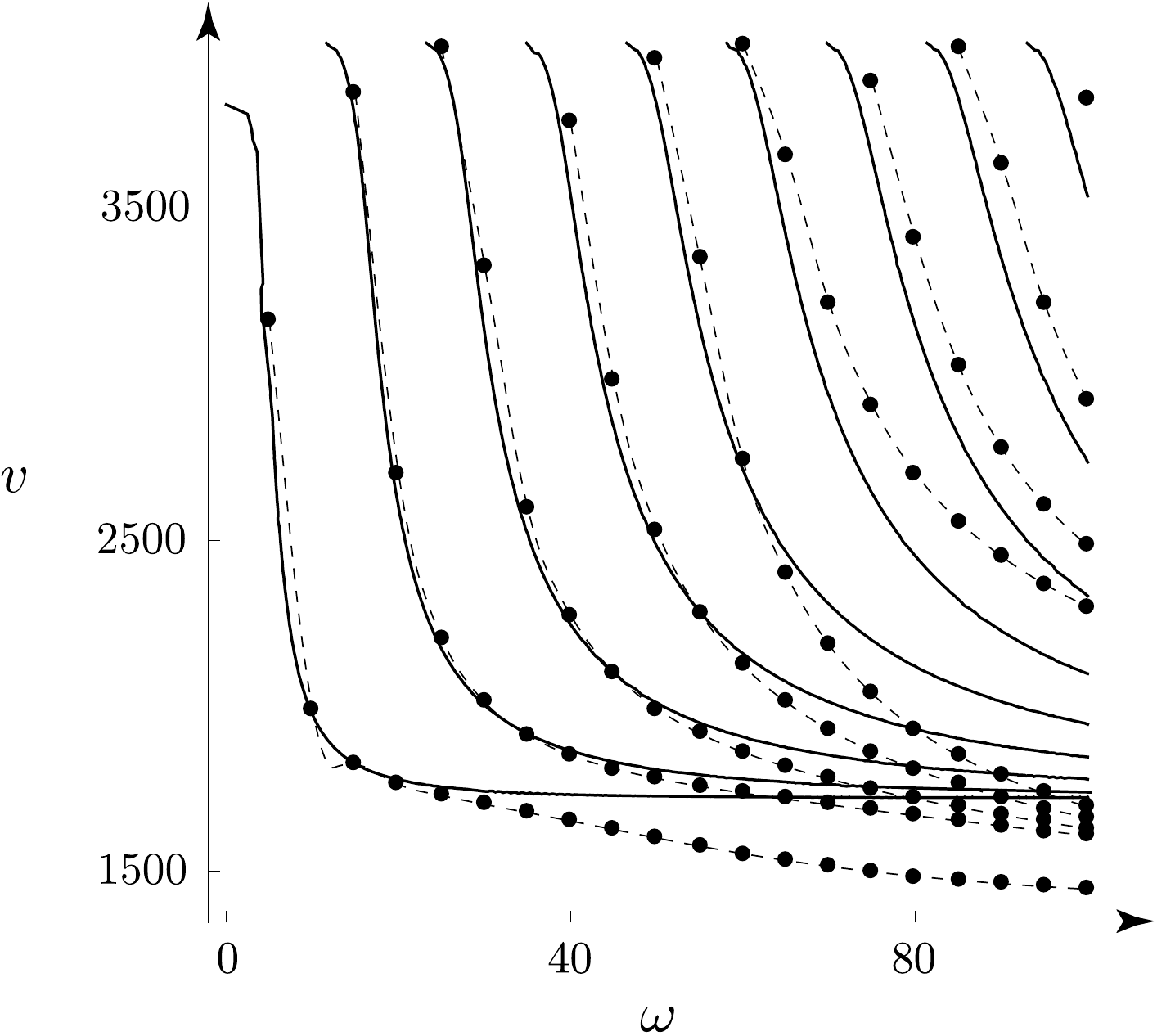}
\end{center}
\caption{\small{Love wave dispersion curves for the Backus medium and delta-matrix solution, shown as black lines and points, respectively}}
\label{fig:loveTI}
\end{figure}

The quasi-Rayleigh wave dispersion curves for the stack of layers given in Table~\ref{table:altTI} are calculated by a Python\textsuperscript{\textregistered} code (Meehan~\cite{meehancode}) based on the reduced-delta-matrix solution of Ikeda and Matsuoka~\cite{ikedamatsuoka}.
We compare these results to those obtained for the Backus medium of that stack by setting to zero the determinant of the matrix in equation~(29) of Khojasteh et al.~\cite{khojasteh}.   
The left-hand plot of Figure~\ref{fig:qRTI} depicts the results for a stack of layers that are five-metres thick and the right-hand plot for a stack of layers that are fifty-metres thick.  
As expected, the match is better for thin layers or low frequencies.

We also calculate Love wave dispersion curves for the same stack of layers using the delta-matrix solution reviewed in Buchen and Ben-Hador~\cite{buchen}, but with the pseudorigidity and pseudothickness defined by Anderson~\cite{anderson62}.
We compare these results to those obtained for the Backus medium of that stack by setting the determinant of the matrix in equation~(30) of Khojasteh et al.~\cite{khojasteh} to zero.   
The left-hand plot of Figure~\ref{fig:loveTI} depicts the results for a stack of layers that are five-metres thick and the right-hand plot for a stack of layers that are fifty-metres thick.  
Again, as expected, the match is better for thin layers or low frequencies.

The left-hand plots of Figures~\ref{fig:qRTI} and \ref{fig:loveTI} correspond to one model; the right-hand plots correspond to another model.
Thus, we can examine the match between the results of the Backus~\cite{backus} average and the propagator matrix for stacks of layers that are five- and fifty-metres thick.
For five-metre layers, the match is good, at all frequencies, for both quasi-Rayleigh and Love waves.
For fifty-metre layers, the match is good for both waves at lower frequencies; at higher frequencies, the match is better for the quasi-Rayleigh wave.
\subsection{Nonalternating layers on transversely isotropic halfspace}
Let us examine the model of twenty transversely isotropic layers overlying a transversely isotropic halfspace used in Harkrider and Anderson~\cite{harkand} and in Ikeda and Matsuoka~\cite{ikedamatsuoka}, to which we refer as {\sl Model HA}.
Let us also examine a modified version of this model, to which we refer as {\sl Model HAS}, where we reduce inhomogeneity.
The mass densities and  elasticity parameters of these models are given in Table~\ref{table:HA}.%
\footnote{In calculating $c_{1133}$\,, we note an error in Anderson~\cite{anderson61}.
In formula for $c_{13}$\,, on page~2955, $(1/2)(c_{11}-c_{33})^2$ should be  $(1/4)(c_{11}-c_{33})^2$
or  $((1/2)(c_{11}-c_{33}))^2$\,.   
It can be confirmed by equations~(9.2.19) and~(9.2.23) of Slawinski~\cite{slawinski1}, with $n_3=\sqrt{2}/2$\,.}
To compute the mean density, we use
\begin{equation}
\label{eq:rho}
\rho^{\overline{\rm TI}}=\overline{\rho}=\frac{1}{n}\sum_{i=1}^n\rho_i
\,.
\end{equation}
For either model, we examine the case of layers whose thickness is five metres and the case of layers whose thickness is one metre.

\begin{figure}
\begin{center}
\includegraphics[scale=0.55]{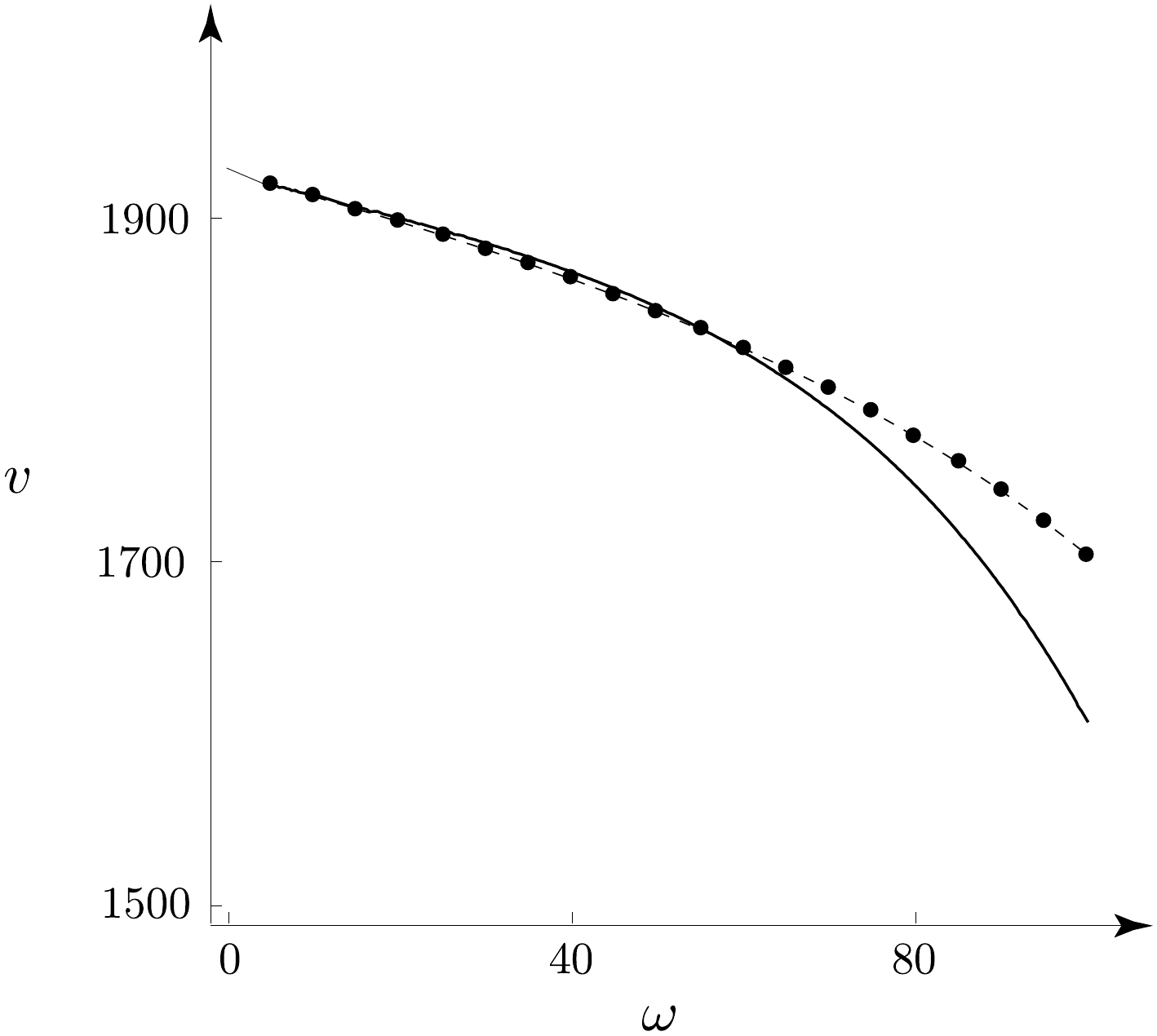}
\includegraphics[scale=0.55]{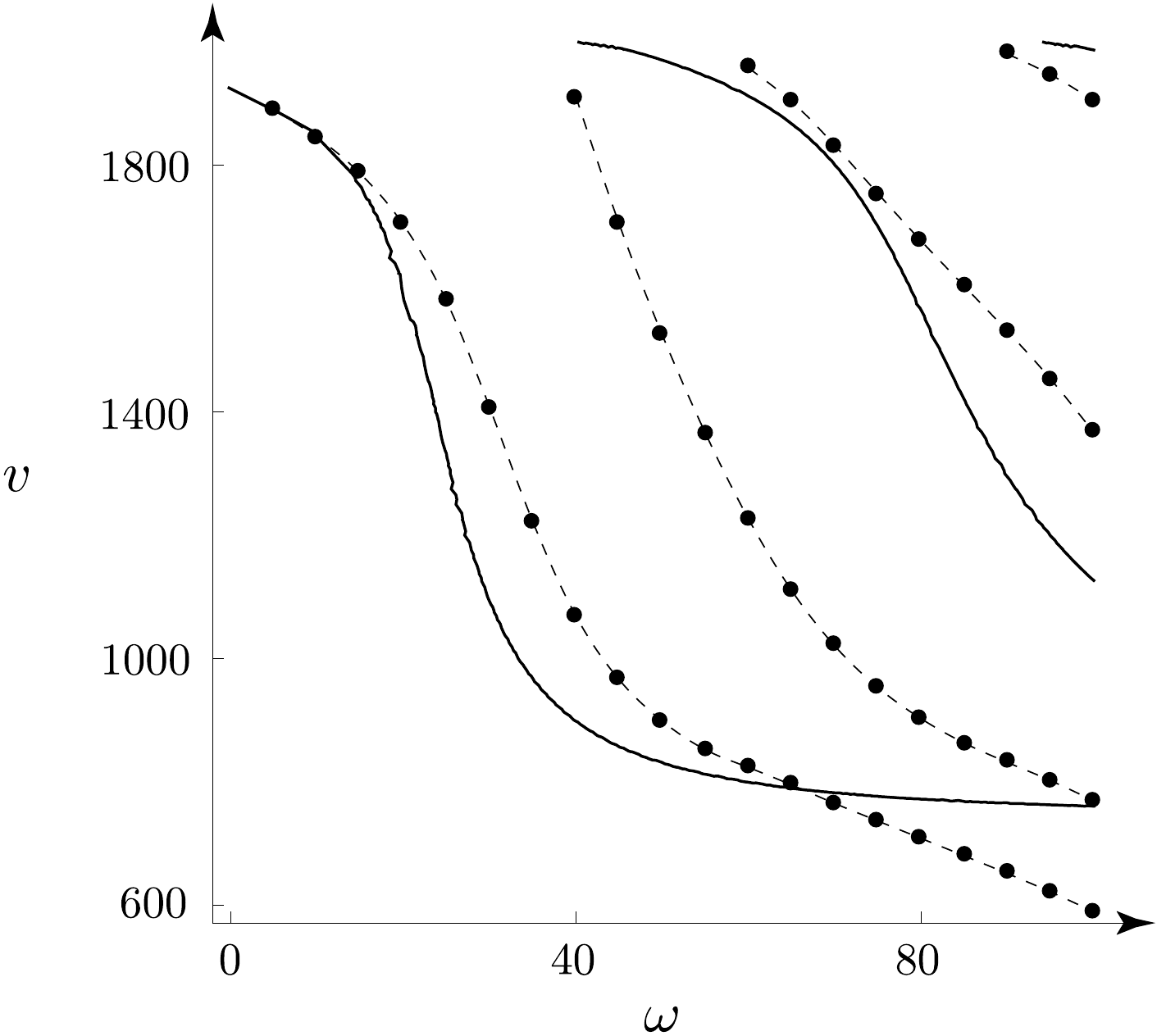}
\end{center}
\caption{\small{quasi-Rayleigh wave dispersion curves for
the Backus medium of {\sl Model HA} and the Ikeda and Matsuoka~\cite{ikedamatsuoka} reduced-delta-matrix solution, shown as black lines and points, respectively}}
\label{fig:qRT-HA}
\end{figure}

In the case of {\sl Model HA}, which has a strong inhomogeneity, we examine Figure~\ref{fig:qRT-HA} to see that---for the case of layers that are one-metre thick, illustrated in the left-hand plot---the results for the Backus medium match the results for the delta-matrix solution for $\omega<70~{\rm s}^{-1}$\,.
This frequency corresponds to the wavelength of about $160~{\rm m}$\,, which is greater than the thickness of the twenty-layer stack and much greater than the thickness of individual layers.
For the case of layers that are five-metres thick, illustrated in the right-hand plot, the results match only for the fundamental mode and for $\omega<15~{\rm s}^{-1}$\,, which corresponds to a wavelength of about $700~{\rm m}$\,.
Again, this wavelength is greater than the thickness of the twenty-layer stack and much greater than the thickness of individual layers.

Backus~\cite{backus} derives the average under the assumption of $\kappa\ell' \ll 1$\,, where $\kappa=\omega/v$ and $\ell'$ is the averaging width, as discussed by Bos et al.~\cite{bos}.
For one-metre layers, $\omega=70~{\rm s}^{-1}$ and
$v=1760~{\rm m}/{\rm s}$\,, we have $\kappa\ell'=0.80$\,. 
For five-metre layers,  $\omega=15~{\rm s}^{-1}$ and $v=1760~{\rm m}/{\rm s}$\,, we have $\kappa\ell'=0.85$\,.
In both cases, the Backus~\cite{backus} average performs better than could be expected in view of its underlying assumption.

\begin{figure}
\begin{center}
\includegraphics[scale=0.55]{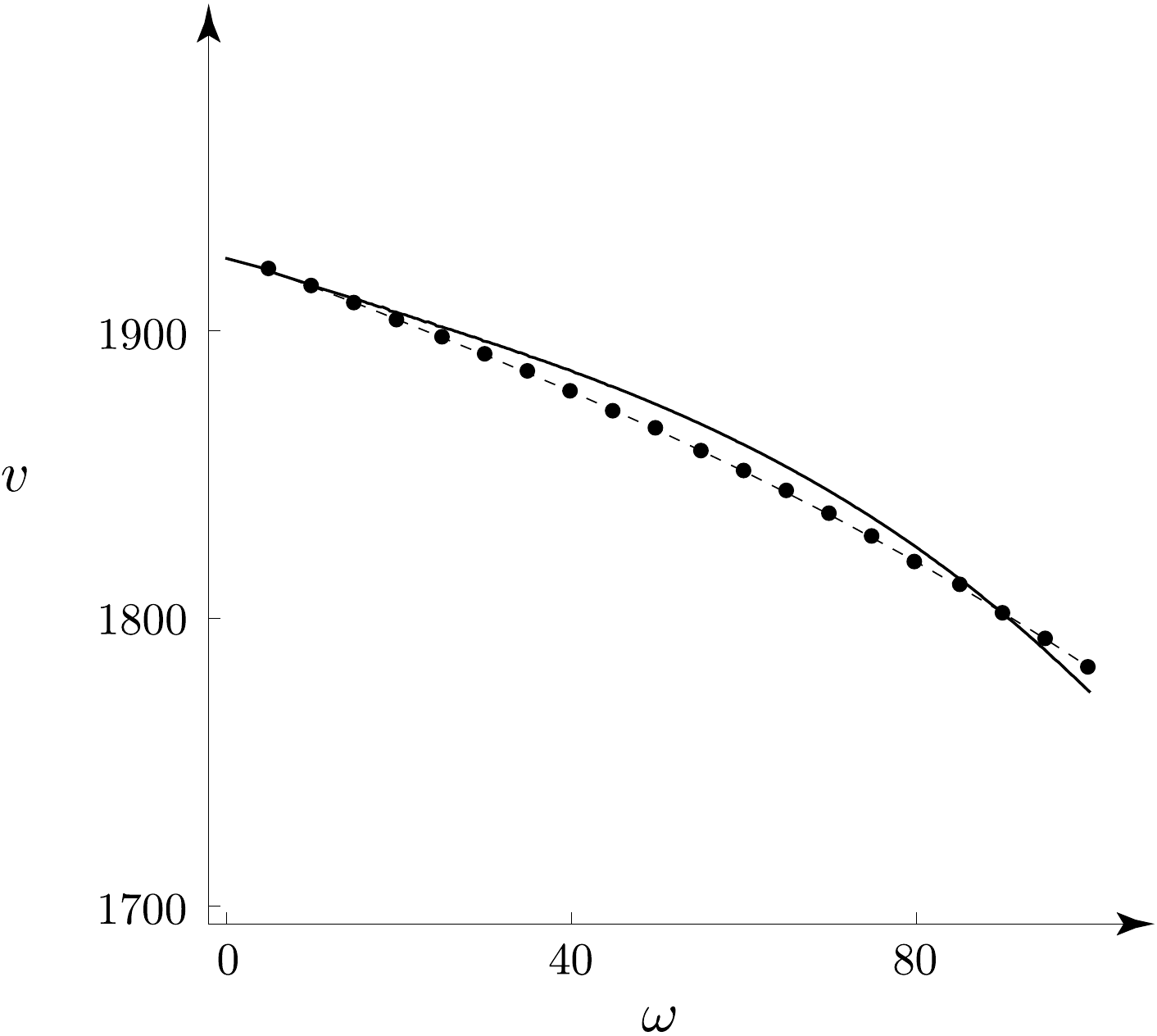}
\includegraphics[scale=0.55]{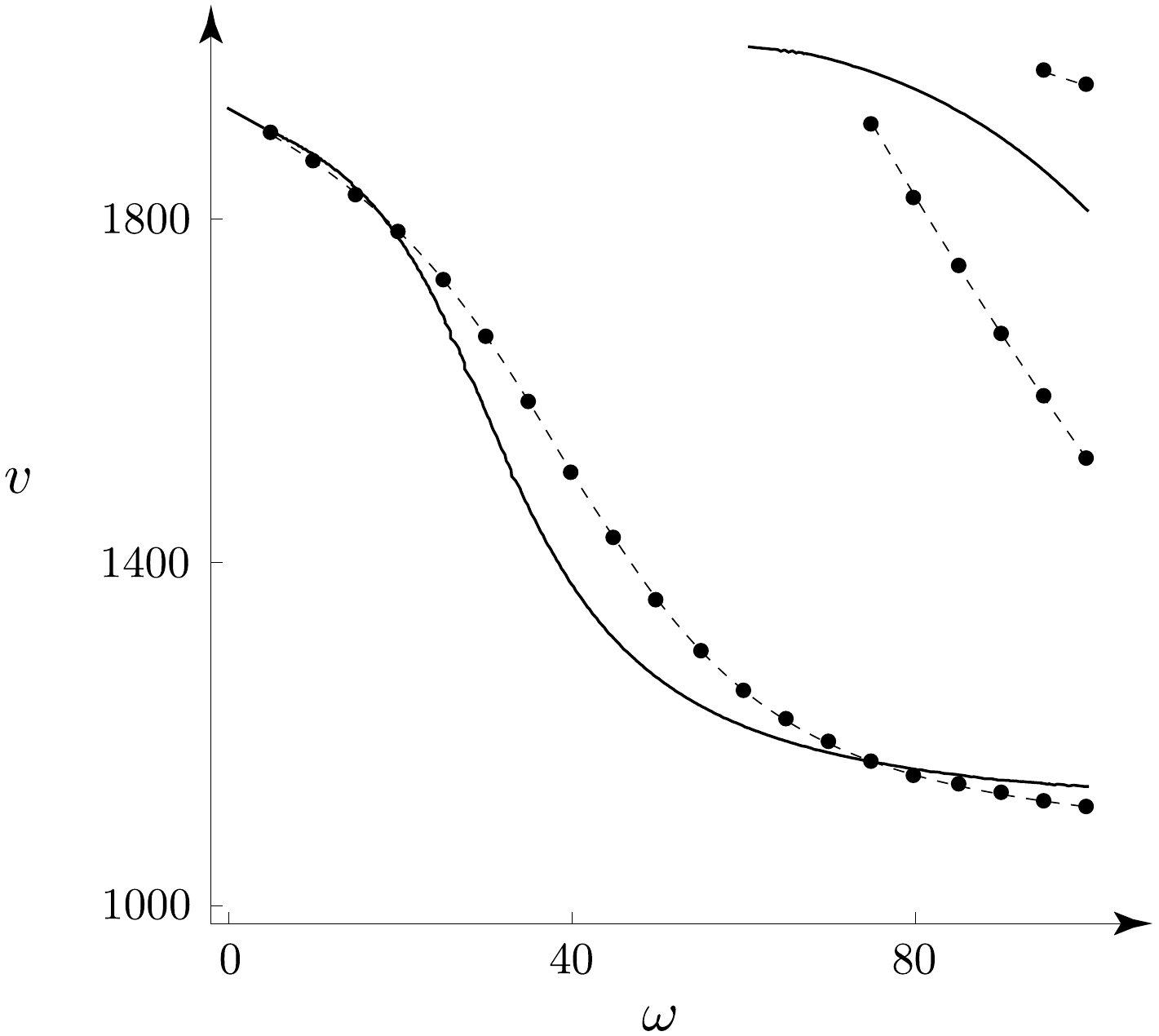}
\end{center}
\caption{\small{quasi-Rayleigh wave dispersion curves for
the Backus medium of {\sl Model HAS} and the Ikeda and Matsuoka~\cite{ikedamatsuoka} reduced-delta-matrix solution, shown as black lines and points, respectively}}
\label{fig:qRT-HAS}
\end{figure}

For {\sl Model HAS}, we examine Figure~\ref{fig:qRT-HAS} to see that, for the case of one-metre layers, illustrated in the left-hand plot, the results for the Backus medium match the results for the delta-matrix solution for $\omega<100~{\rm s}^{-1}$\,.
For the case of five-metre layers, illustrated in the right-hand plot, the match is good only for $\omega<20~{\rm s}^{-1}$\,, since, beyond this point, the discrepancy of values for~$v$ between two steep curves is significant.

\begin{table}
\begin{center}
\begin{tabular}{|c||c|c|c|c|c|c|}
\hline
layer & $\rho$ & $c_{1111}$ & $c_{1133}$ & $c_{3333}$ & $c_{2323}$ & $a$ \\
\hline\hline
1 & 2000 & 2.90 & 2.63 & 2.90 & 0.14 & 4.00\\
2 & 2000 & 4.43 & 3.62 & 4.43 & 0.40 & 1.90\\
3 & 2000 & 3.88 & 1.74 & 3.88 & 1.07 & 1.85\\
4 & 2250 & 5.80 & 1.30 & 5.80 & 2.25 & 1.80\\
5 & 2250 & 6.50 & 8.37 & 5.89 & 2.65 & 1.75\\
6 & 2250 & 6.50 & 1.06 & 5.89 & 2.54 & 1.70\\
7 & 2250 & 7.25 & 1.64 & 6.57 & 2.54 & 1.65\\
8 & 2250 & 7.25 & 1.23 & 6.57 & 2.01 & 1.60\\
9 & 2250 & 7.83 & 2.04 & 4.92 & 2.25 & 1.55\\
10 & 2250 & 7.83 & 1.44 & 4.92 & 2.54 & 1.50\\
11 & 2250 & 9.00 & -4.27 & 5.65 & 4.52 & 1.45\\
12 & 2250 & 11.33 & -1.98 & 7.12 & 4.52 & 1.40\\
13 & 2500 & 37.97 & 20.69 & 36.07 & 6.45 & 1.35\\
14 & 2500 & 73.12 & 51.02 & 69.46 & 6.83 & 1.30\\
15 & 2500 & 73.12 & 54.84 & 62.15 & 6.64 & 1.25\\
16 & 2500 & 73.12 & 54.84 & 62.15 & 6.64 & 1.20\\
17 & 2500 & 81.00 & 49.94 & 68.85 & 10.31 & 1.15\\
18 & 2500 & 78.32 & 52.56 & 66.57 & 7.84 & 1.10\\
19 & 2500 & 75.71 & 51.81 & 68.14 & 8.48 & 1.05\\
20 & 2500 & 93.54 & 70.12 & 84.19 & 7.43 & 1.00\\
\hline
 H & 2600 & 101.08 & 71.02 & 90.97& 10.40 & 1.00\\
\hline
\end{tabular}
\end{center}
\caption{\small{Mass densities, in ${\rm kg}/{\rm m}^{3}$\,,
 and elasticity parameters, in $10^9\,{\rm N}/{\rm m}^{2}$\,, of {\sl Model HA}\,; elasticity parameters of {\sl Model HAS} are obtained by multiplying elasticity parameters of {\sl Model HA} by factor~$a$\,.}}
\label{table:HA}
\end{table}

 {\sl Model HA} and {\sl Model HAS} do not include values of $c_{1212}$ or $c_{1122}$\,, so we are unable to generate their Love-wave dispersion curves, which would be the counterparts of Figures~\ref{fig:qRT-HA} and \ref{fig:qRT-HAS}.
 Thus, herein, we cannot compare the behaviours of the quasi-Rayleigh and Love waves for nonalternating layers overlying a transversely isotropic halfspace.
\section{Conclusion}
Comparing the modelling of guided waves using the Backus~\cite{backus} average to modelling of these waves based on the propagator matrix, we obtain a good match for the fundamental mode in weak inhomogeneity of layers and, as expected, for low frequencies or thin layers.

For alternating layers, which is a common occurrence in sedimentary basins, the discrepancy between these two methods remains small, even for strong inhomogeneity. 
For higher modes, the difference remains small near the cutoff frequency.  

For a stack of nonalternating transversely isotropic layers that is strongly inhomogeneous, the discrepancy is small only for the fundamental mode and for low frequencies or thin layers.
The results become similar---even for higher frequencies---if the layers become thinner and if the inhomogeneity is diminished.
Also, in such a case, similar results are obtained for $\kappa\ell'\lesssim 1$\,, in spite of the underlying assumption,~$\kappa\ell'\ll 1$\,.   

Let us comment on the results, in the context of conclusions presented in previous works.
Liner and Fei~\cite{linerfei} recommend the averaging length be less than or equal to one-third of the dominant seismic wavelength, which corresponds to $\kappa\ell'\leqslant 2$\,.
This is the point at which the fundamental mode solutions shown in the right-hand plot of Figure~\ref{fig:CTI100-R} begin to diverge from one another.
However, upon examining Figures~\ref{fig:CTI500-R}--\ref{fig:qRT-HAS}, we conclude that we do not have a single maximum value of $\kappa\ell'$ at which the solutions begin to diverge; it depends on the context.
For instance, for the fundamental mode, the maximum value of $\kappa\ell'$ ranges from $0.80$\,, as is the case in the left-hand plot of Figure~\ref{fig:qRT-HA}, to $20$\,, in the right-hand plot of Figure~\ref{fig:qRTI}.
Yet, for Figures~\ref{fig:CTI500-R}--\ref{fig:qRT-HAS}, the median value of $\kappa\ell'$ at which the solutions begin to diverge is $2$\,, which is the value suggested by Liner and Fei~\cite{linerfei}.

Mavko et al.~\cite{mavko} suggest the necessity for layers to be at least ten times smaller than the seismic wavelength,~$\lambda/h>10$\,, where $\lambda=2\pi v/\omega$\,, with $v$ standing for the propagation speed of the quasi-Rayleigh or Love wave, and $h$ is the layer thickness.
Again, examining Figures~\ref{fig:CTI500-R}--\ref{fig:qRT-HAS}, we conclude that we do not have a single  minimum $\lambda/h$ ratio, at which the solutions begin to diverge.
For instance, for the fundamental mode, the minimum value of $\lambda/h$ ranges from $3$\,, in the right-hand plot of Figure~\ref{fig:qRTI}\,, to $150$\,, in the left-hand plot of Figure~\ref{fig:qRT-HA}.
For Figures~\ref{fig:CTI500-R}--\ref{fig:qRT-HAS}, the median value of $\lambda/h$ at which the solutions begin to diverge is $28$\,, which is of the same order of magnitude as the value suggested by Mavko et al.~\cite{mavko}.

Capdeville et al.~\cite{cap2013} state that the Backus~\cite{backus} average is applicable only to a fine-scale layered medium, far from the free surface and from the source.
Our results are consistent with the first part of that statement.
However, in contrast to the middle part of that statement, for most cases, we obtain satisfactory results, even in proximity of the free surface.
Results are degraded for {\sl Model HA} due to near-surface low-velocity layers. 
 We cannot examine the last part of that statement, which is the issue of source proximity.

In comparing the results obtained for a Backus medium to the results obtained for a Voigt medium, we can treat the latter as an approximation of the former.
Both are viewed as analogies for the behaviour of seismic waves in thinly layered media.
It is not the case in comparing the results obtained for a Backus medium to the results obtained with a propagator matrix.
The latter is not restricted to the assumption of~$\kappa\ell'\ll 1$\,.
This restriction on the Backus~\cite{backus} average is essential to its importance in modelling seismic response of thinly layered media.
The purpose of this average is not to provide information about the material itself, but to model its response to a seismic signal.
As such, the frequency where the discrepancy between the results obtained for a Backus medium and the results obtained with a propagator matrix becomes significant can be interpreted as the frequency beyond which the Backus~\cite{backus} average ceases to be empirically adequate in the context of seismology.
The adequacy of the propagator matrix depends on the frequency content of a seismic signal.

Results of this study, in particular empirical adequacy of modelling techniques, allow us to gain insight into the reliability of a joint inverse of the quasi-Rayleigh and Love dispersion curves for obtaining model parameters.
A work on that subject is presented by Bogacz et al.~\cite{bogacz}.
\section*{Acknowledgments}
We wish to acknowledge discussions with Md Abu Sayed, Piotr Stachura, and Theodore Stanoev, email communication with Tatsunori Ikeda, the graphic support of Elena Patarini, and proofreading of Theodore Stanoev.
This research was performed in the context of The Geomechanics Project supported by Husky Energy.
Also, this research was partially supported by the Natural Sciences and Engineering Research Council of Canada, grant 238416-2013.
\bibliographystyle{unsrt}
\bibliography{DMS}

\begin{thebibliography}{10}

\bibitem{backus}
G.~E. Backus.
\newblock Long-wave elastic anisotropy produced by horizontal layering.
\newblock {\em J. Geophys. Res.}, 67(11):4427--4440, 1962.

\bibitem{voigt}
W.~Voigt.
\newblock {\em {L}ehrbuch der {K}ristallphysik}.
\newblock Teubner, Leipzig, 1910.

\bibitem{dalton}
D.~R. Dalton, M.~A. Slawinski, P.~Stachura, and T.~Stanoev.
\newblock Sensitivity of {L}ove and quasi-{R}ayleigh waves to model parameters.
\newblock {\em Q. J. Mech. Appl. Math.}, 70(2):103--130, 2017.

\bibitem{anderson62}
D.~L. Anderson.
\newblock Love wave dispersion in heterogeneous anisotropic media.
\newblock {\em Geophysics}, 27(4):445--454, 1962.

\bibitem{postma}
G.~W. Postma.
\newblock Wave propagation in a stratified medium.
\newblock {\em Geophysics}, 20:780--806, 1955.

\bibitem{thomsen}
L.~Thomsen.
\newblock Weak elastic anisotropy.
\newblock {\em Geophysics}, 51(10):1954--1966, 1986.

\bibitem{bos}
L.~Bos, D.~R. Dalton, M.~A. Slawinski, and T.~Stanoev.
\newblock On {B}ackus average for generally anisotropic layers.
\newblock {\em Journal of Elasticity}, 127(2):179--196, 2017.

\bibitem{BosX}
L.~Bos, T.~Danek, M.~A. Slawinski, and T.~Stanoev.
\newblock Statistical and numerical considerations of {B}ackus-average product
  approximation.
\newblock {\em Journal of Elasticity}, DOI 10.1007/s10659-017-9659-9, 2017.

\bibitem{slawinski3}
M.~A. Slawinski.
\newblock {\em Waves and rays in seismology: {A}nswers to unasked questions}.
\newblock World Scientific, 2016.

\bibitem{gazis}
D.~C. Gazis, I.~Tadjbakhsh, and R.~A. Toupin.
\newblock The elastic tensor of given symmetry nearest to an anisotropic
  elastic tensor.
\newblock {\em Acta Crystallographica}, 16(9):917--922, 1963.

\bibitem{Udias1999}
A.~Ud\'{i}as.
\newblock {\em Principles of seismology}.
\newblock Cambridge University Press, 1999.

\bibitem{khojasteh}
A.~Khojasteh, M.~Rahimian, R.~Y.~S. Pak, and M.~Eskandari.
\newblock Asymmetric dynamic {G}reen's functions in a two-layered transversely
  isotropic half-space.
\newblock {\em J. Eng. Mech.}, 134(9):777--787, 2008.

\bibitem{bagheri}
A.~Bagheri, S.~Greenhalgh, A.~Khojasteh, and M.~Rahimian.
\newblock Dispersion of {R}ayleigh, {S}cholte, {S}toneley and {L}ove waves in a
  model consisting of a liquid layer overlying a two-layer transversely
  isotropic solid medium.
\newblock {\em Geophys. J. Int.}, 203:195--212, 2015.

\bibitem{buchen}
P.~W. Buchen and R.~Ben-{H}ador.
\newblock Free-mode surface-wave computations.
\newblock {\em Geophys. J. Int.}, 124:869--887, 1996.

\bibitem{meehancode}
T.~B. Meehan.
\newblock Python\textsuperscript{\textregistered} code for calculating surface
  wave dispersion curves.\newline
\newblock https://github.com/tbmcoding/dispersion, December 2017.

\bibitem{meehanarxiv}
T.~B. Meehan.
\newblock Evolution of the propagator matrix method and its implementation in
  seismology.
\newblock {\em ar{X}iv}, 1801.04635([physics.geo-ph]):1--12, January 2018.

\bibitem{ikedamatsuoka}
T.~Ikeda and T.~Matsuoka.
\newblock Computation of {R}ayleigh waves on transversely isotropic media by
  the reduced delta matrix method.
\newblock {\em Bulletin of the Seismological Society of America},
  103(3):2083--2093, 2013.

\bibitem{brisco}
C.~Brisco.
\newblock Anisotropy vs. inhomogeneity: Algorithm formulation, coding and
  modelling.
\newblock Honours thesis, Memorial University, 2014.

\bibitem{DNS}
T.~Danek, A.~Noseworthy, and M.~A. Slawinski.
\newblock Effects of norms on general {H}ookean solids for their isotropic
  counterparts.
\newblock {\em Dolomites research notes on approximation}, 11:1--14, 2018
  (accepted).

\bibitem{harkand}
D.~G. Harkrider and D.~L. Anderson.
\newblock Computation of surface wave dispersion for multilayered anisotropic
  media.
\newblock {\em Bulletin of the Seismological Society of America},
  52(2):321--332, 1962.

\bibitem{anderson61}
D.~L. Anderson.
\newblock Elastic wave propagation in layered anisotropic media.
\newblock {\em Journal of Geophysical Research}, 66(9):2953--2963, 1961.

\bibitem{slawinski1}
M.~A. Slawinski.
\newblock {\em Waves and rays in elastic continua}.
\newblock World Scientific, Singapore, 3rd edition, 2015.

\bibitem{linerfei}
C.~L. Liner and T.~W. Fei.
\newblock Layer-induced seismic anisotropy from full-wave sonic logs: theory,
  applications and validation.
\newblock {\em Geophysics}, 71:D183--D190, 2006.

\bibitem{mavko}
G.~Mavko, T.~Mukerji, and J.~Dvorkin.
\newblock {\em The Rock Physics Handbook}.
\newblock Cambridge University Press, Cambridge, 1998.

\bibitem{cap2013}
Y.~Capdeville, \'E. Stutzmann, N.~Wang, and {J.-P.} Montagner.
\newblock Residual homogenization for seismic forward and inverse problems in
  layered media.
\newblock {\em Geophys. J. Int.}, 194:470--487, 2013.

\bibitem{bogacz}
A.~Bogacz, D.~R. Dalton, T.~Danek, K.~Miernik, and M.~A. Slawinski.
\newblock On {P}areto {J}oint {I}nversion of guided waves.
\newblock {\em ar{X}iv}, 1712.09850:[physics.geo--ph], 2017.

\end{thebibliography}
\end{document}